\documentclass[12pt]{article}
\usepackage{amsthm,amsfonts,amsmath,amssymb,mathrsfs,graphicx,url,tikz,hyperref}

\title{Positron Position Operators. I.\\ A Natural Option}

\author{
Roderich Tumulka\footnote{Fachbereich Mathematik, Eberhard-Karls-Universit\"at, Auf der Morgenstelle 10, 72076 T\"ubingen, Germany. E-mail: roderich.tumulka@uni-tuebingen.de}
}
\date{November 23, 2021}

\newcommand{\be}{\begin{equation}}
\newcommand{\ee}{\end{equation}}
\newcommand{\Hilbert}{\mathscr{H}}
\newcommand{\Kilbert}{\mathscr{K}}
\newcommand{\conf}{\mathcal{Q}}
\newcommand{\Q}{\mathcal{Q}}
\newcommand{\lattice}{\mathcal{L}}
\newcommand{\sL}{\mathscr{L}}
\newcommand{\Fock}{\mathscr{F}}
\newcommand{\alg}{\mathcal{A}}
\newcommand{\open}{\mathcal{O}}

\renewcommand{\Im}{\mathrm{Im}}
\newcommand{\scp}[2]{\langle #1|#2 \rangle}

\newcommand{\pr}[1]{| #1 \rangle \langle #1 |}

\newcommand{\obv}{{\mathrm{obv}}}
\newcommand{\nat}{{\mathrm{nat}}}
\newcommand{\pre}{{\mathrm{pre}}}
\newcommand{\lf}{{\mathrm{loc.fin}}}

\newcommand{\Anti}{{\mathrm{Anti}}}

\newcommand{\el}{{\mathrm{el}}}
\newcommand{\pos}{{\mathrm{pos}}}
\newcommand{\ket}[1]{|#1\rangle}

\newcommand{\CCC}{\mathbb{C}}

\newcommand{\MMM}{\mathbb{M}}
\newcommand{\NNN}{\mathbb{N}}
\newcommand{\PPP}{\mathbb{P}}
\newcommand{\RRR}{\mathbb{R}}

\newcommand{\TTT}{\mathbb{T}}
\newcommand{\ZZZ}{\mathbb{Z}}

\newcommand{\va}{\boldsymbol{a}}
\newcommand{\ve}{\boldsymbol{e}}

\newcommand{\vy}{\boldsymbol{y}}
\newcommand{\valpha}{\boldsymbol{\alpha}}
\newcommand{\vx}{\boldsymbol{x}}
\newcommand{\vxbar}{\overline{\vx}}
\newcommand{\vX}{\boldsymbol{X}}

\newcommand{\jbar}{{\overline{\jmath}}}
\newcommand{\nbar}{{\overline{n}}}

\newcommand{\sbar}{{\overline{s}}}
\newcommand{\xbar}{{\overline{x}}}
\newcommand{\mubar}{{\overline{\mu}}}
\DeclareMathOperator{\tr}{tr}

\newcommand{\sC}{\mathscr{C}}
\newcommand{\sD}{\mathscr{D}}

\newcommand{\sM}{\mathscr{M}}

\theoremstyle{theorem}

\newtheorem{conj}{Conjecture}
\theoremstyle{definition}
\newtheorem{defn}{Definition}

\begin{document}

\maketitle

\begin{abstract}
By ``position operators,'' I mean here a POVM (positive-operator-valued measure) on a suitable configuration space acting on a suitable Hilbert space that serves as defining the position observable of a quantum theory, and by ``positron position operators,'' I mean a joint treatment of positrons and electrons. I consider the standard free second-quantized Dirac field in Minkowski space-time or in a box. On the associated Fock space (i.e., the tensor product of the positron Fock space and the electron Fock space), there acts an obvious POVM $P_\obv$, but I propose a different one that I call the natural POVM, $P_\nat$. In fact, it is a PVM (projection-valued measure); it captures the sense of locality corresponding to the field operators $\Psi_s(\vx)$ and to the algebra of local observables. The existence of $P_\nat$ depends on a mathematical conjecture which at present I can neither prove nor disprove; here I explore consequences of the conjecture. I put up for consideration the possibility that $P_\nat$, and not $P_\obv$, is the physically correct position observable and defines the Born rule for the joint distribution of electron and positron positions. I describe properties of $P_\nat$, including a strict no-superluminal-signaling property, and how it avoids the Hegerfeldt-Malament no-go theorem. I also point out how to define Bohmian trajectories that fit together with $P_\nat$, and how to generalize $P_\nat$ to curved space-time.

\medskip

\end{abstract}

\newpage

\tableofcontents

\section{Introduction}
\label{sec:intro}

In the standard representation of vectors from the Hilbert space of the free Dirac quantum field, they can be written as wave functions $\psi$ of the positions of a variable number of particles; the associated $|\psi|^2$ distribution over particle position configurations corresponds to position operators that I call the obvious position operators. In this paper I suggest that the obvious position operators are not physically correct and introduce instead a different set of position operators that is more closely linked to the position variable $\vx$ in the field operators $\Psi_s(\vx)$ and to what is called the local observables in algebraic quantum field theory.

I begin by making explicit the obvious position operators, then I will say something about why it is desirable to have position operators at all, and then, in Sections~\ref{sec:Pnatdef}--\ref{sec:last}, I will introduce and discuss the new proposal. I discuss another possibility in \cite{positron2}.

\subsection{The Obvious Position Operators}
\label{sec:obv}

The standard free quantized Dirac field on Minkowski space-time is defined (e.g., \cite{Tha}) on a Hilbert space $\Hilbert$ that is the tensor product of two Fock spaces, one for the electrons and one for the positrons, 
\be
\Hilbert= \Fock(\Hilbert_{1+}) \otimes \Fock(\overline{\Hilbert_{1-}}).
\ee
Here, $\Fock(\Kilbert)$ means the fermionic Fock space of the 1-particle Hilbert space $\Kilbert$,
\be
\Fock(\Kilbert) = \bigoplus_{n=0}^\infty \Anti \,\Kilbert^{\otimes n}\,,
\ee
where $\Anti$ denotes the anti-symmetrization operator; $\Hilbert_{1\pm}$ means the positive (negative) spectral subspace of the 1-particle Dirac operator ($\hbar=1=c$)
\be
H_1 = -i\valpha \cdot \nabla + m\beta
\ee
acting in the 1-particle Hilbert space
\be
\Hilbert_1 = L^2(\RRR^3,\CCC^4);
\ee
and $\overline{\Kilbert}$ means the complex conjugate space of the Hilbert space $\Kilbert$ (see Appendix~\ref{app:conjugate}). Since every element of $\Hilbert_1$ can be expressed as a function
\be
\psi_s(\vx)
\ee
with $\vx$ running through $\RRR^3$ and the index $s\in\{1,2,3,4\}$ labeling the 4 dimensions of Dirac spin space, and since $\overline{\Hilbert_{1-}}$ can be canonically and naturally identified with $\Hilbert_{1+}$ by means of the anti-unitary charge conjugation operator $C$, 
\be
C\psi(\vx) = i\alpha_2 \psi^*(\vx)\,,
\ee
every element $\psi$ of $\Hilbert$ can be expressed as a function on the configuration space
\be\label{confhatdef}
\widehat\conf = \widehat\conf_{\el} \times \widehat\conf_{\pos} = \biggl(\bigcup_{n=0}^\infty (\RRR^3)^n \biggr) \times \biggl( \bigcup_{\nbar=0}^\infty (\RRR^3)^{\nbar}\biggr) = \bigcup_{n,\nbar=0}^\infty \underbrace{(\RRR^3)^{n+\nbar}}_{=:\widehat\conf^{n\nbar}}
\ee
(where $\nbar$ is the name of an integer-valued index and does not involve conjugation), which we will write as
\be
\psi_{s_1...s_n\sbar_1...\sbar_\nbar}(\vx_1,\ldots,\vx_n,\vxbar_1,\ldots,\vxbar_{\nbar})
\ee
(where $\vxbar$ is the name of a variable running through $\RRR^3$).

Therefore, there is an obvious candidate for the probability density $\rho$ of the positions of the electrons and positrons in the quantum state $\psi\in\Hilbert$ with $\|\psi\|=1$:
\be
\rho_{\obv}=|\psi|^2\,,
\ee
that is,
\be
\rho_{\obv}(\vx_1...\vx_n,\vxbar_1...\vxbar_{\nbar}) = 
\sum_{s_1...s_n,\sbar_1...\sbar_{\nbar}=1}^4 \Bigl|\psi_{s_1...s_n\sbar_1...\sbar_\nbar}(\vx_1...\vx_n,\vxbar_1...\vxbar_{\nbar}) \Bigr|^2
\ee
for all $(\vx_1...\vx_n,\vxbar_1...\vxbar_{\nbar})\in\widehat\conf$.
It is normalized in the sense that
\begin{align}
1&=\int_{\conf}dq \: \rho_{\obv}(q)\\
&=\sum_{n,\nbar=0}^\infty \int\limits_{\RRR^3}d^3\vx_1\cdots\int\limits_{\RRR^3}d^3\vx_n\int\limits_{\RRR^3}d^3\vxbar_1\cdots \int\limits_{\RRR^3}d^3\vxbar_{\nbar} ~ \rho_{\obv}(\vx_1...\vx_n,\vxbar_1...\vxbar_{\nbar})\,.
\end{align}

In the following, it will be useful to distinguish between \emph{ordered} and \emph{unordered} configurations:  An ordered configuration of $n$ particles at positions $\vx_1,\ldots,\vx_n$ is the $n$-tuple $(\vx_1,\ldots,\vx_n)$, whereas an unordered configuration of $n$ particles at the same positions is the set $\{\vx_1,\ldots,\vx_n\}$. It will be convenient to write wave functions $\psi$ as functions on ordered configuration spaces such as $\widehat\conf$ as in \eqref{confhatdef} but define measures on unordered configuration spaces such as
\be\label{confdef}
\conf = \conf_\el \times \conf_\pos = \Gamma(\RRR^3) \times \Gamma(\RRR^3)
\ee
with
\be
\Gamma(S) := \{q\subset S: \# q<\infty\}
\ee
the set of finite subsets.
The ``unordering map'' $\pi:\widehat\conf\to\conf$ is defined by
\be\label{unorderingdef}
\pi(\vx_1...\vx_n,\vxbar_1...\vxbar_\nbar)
=(\{\vx_1...\vx_n\},\{\vxbar_1...\vxbar_\nbar\})\,.
\ee

So the density function $\rho_\obv:\widehat\conf\to [0,\infty)$ defines a probability measure $\PPP_\obv$ on $\conf$ according to\footnote{Here and in the following, I simply refer to ``all subsets'' when I mean ``all measurable subsets.''}
\be
\PPP_\obv(B) := \int\limits_{\pi^{-1}(B)} \!\! dq \: \rho_\obv(q)~~~\forall B \subseteq \conf \,,
\ee
where $dq$ refers to volume in $\widehat\conf$. The measure $\PPP_\obv$ can be expressed in terms of a POVM $P_\obv$ on $\conf$ acting on $\Hilbert$ (see Appendix~\ref{app:POVM} for details),
\be
\PPP_{\obv}(B) = \scp{\psi}{P_\obv(B)|\psi} ~~~\forall B \subseteq \conf\,.
\ee
I refer to the POVM as ``the position operators.'' In general, $P_\obv(B)$ does not commute with $P_\obv(B')$.

In this paper, I introduce a different POVM that I call the natural POVM $P_\nat$, and I present reasons for thinking that the corresponding probability measure $\PPP_\nat$ defined by
\be
\PPP_\nat(B) = \scp{\psi}{P_\nat(B)|\psi} ~~~\forall B \subseteq \conf
\ee
is the physically correct joint probability distribution of the positions of the positrons and electrons (so that $P_\nat$ provides the physically correct position operators of the positrons and electrons). 
The considerations here are non-rigorous; even the existence of $P_\nat$ depends on a mathematical conjecture which I can neither prove nor disprove. 

The vector
\be
|\Omega\rangle = (1,0,0,0,\ldots)\otimes (1,0,0,0,\ldots) \in \Hilbert
\ee
is usually called the ``vacuum state'' but I will call it the ``sea state'' because, although it corresponds to the empty configuration in terms of $P_\obv$,
\be
P_\obv(\{\emptyset\}) = \pr{\Omega}\,,
\ee
this is not the case for $P_\nat$. However, it is still the case that $|\Omega\rangle$ is the ground state (state of lowest energy) of the free Hamiltonian on $\Hilbert$.

\subsection{What Position Operators Are Good For}

It is widespread practice in quantum electrodynamics not to consider any probability distribution over positron or electron positions; instead of position operators, one considers only field operators. But even in this widespread spirit it will be of interest to know that there is a POVM better suited as the position observable than $P_\obv$---it does not hurt to have a position POVM. Even more, readers may want to reconsider the status of position configuration space and position probability distributions in the light of the new theoretical possibilities opened up by $P_\nat$. In particular, so far it may have seemed that the Hegerfeldt-Malament theorem \cite{Heg74,Mal96,HC02} or related results exclude the possibility of a convincing position POVM; I will discuss in Section~\ref{sec:Malament} why these results do not apply to $P_\nat$.

Position operators can appear in quantum theories in three roles: First, as representing the position observable in the sense of defining the probability distribution of the configuration that detectors will find. Second, as defining the distribution that actual particle positions have, regardless of observation, in interpretations of quantum theory (such as Bohm's \cite{DT09}) in which particles have actual positions. And third, as defining, from given vectors in $\Hilbert$, the probabilities of macroscopic situations such as a cat being alive or the needle of a measurement apparatus pointing to a particular value; without position operators, or rather without a Born rule asserting position probabilities, the logical link between quantum states of cats or apparatus and probabilities of their macroscopic configuration would be missing \cite{Beck}.\footnote{A different strategy for providing this missing link by reducing the question to measurements at $t=\infty$ has been proposed by Fewster and Verch \cite{FV18,Few19}.} As will become clear, $P_\nat$ is suitable for all three roles.

It is also widespread practice to focus on $t=\pm\infty$, in part because the time evolution from $t=-\infty$ to $t=+\infty$ (the scattering matrix) is often better or more easily defined than that between finite times, and also in part because position distributions are often better or more easily defined at $t=\pm\infty$ than at intermediate times. Specifically, as $t\to\pm\infty$, wave functions tend to become locally plane waves, and contributions with different momenta tend to separate in position, so that the position distribution at $t=\infty$ agrees with the momentum distribution, which is easily defined. As a consequence, to specify a position distribution also helps us deal more easily with finite times.

\bigskip

The remainder of this paper is organized as follows. In Section~\ref{sec:Pnatdef}, I describe the definition of $P_\nat$. In Section~\ref{sec:Pnatfinite}, I focus on the simpler case of particles in a box. In Section~\ref{sec:discrete}, I describe an alternative approach to $P_\nat$ based on discretizing space and taking a continuum limit later. In Section~\ref{sec:Diracsea}, I argue that the idea of taking the Dirac sea literally, advocated in particular by \cite{BH,CS07,DEO17}, should, after suitable mathematical consideration, lead as well to $P_\nat$. In Section~\ref{sec:Pobv}, I describe arguments against $P_\obv$. In Section~\ref{sec:properties}, I discuss properties of $P_\nat$ and explain how it avoids known no-go results such as the Hegerfeldt-Malament theorem and the Reeh-Schlieder theorem. In Section~\ref{sec:curved}, I discuss how $P_\nat$ can be generalized to curved space-time. In Section~\ref{sec:last}, I conclude.

\section{Definition of $P_\nat$}
\label{sec:Pnatdef}

The definition of $P_\nat$ is based on charge operators $Q(A)$ associated with regions $A$ in 3-space, which are in turn defined in terms of the field operators $\Psi_s(\vx)$. To make their definition explicit, we abbreviate $(\vx,s)$ by $x$ and recall the standard definition \cite{Tha} of the electron/positron annihilation/creation operators: for $f\in \Hilbert_{1+}$ and $g\in\Hilbert_{1-}$,
\begin{subequations}\label{abdef}
\begin{align}
a(f)\psi^{n\nbar}(x_1...x_n, \xbar_1...\xbar_\nbar)
&=\sqrt{n+1}\sum_{s=1}^4\int\!\! d^3\vx\: f(x)^* \: \psi^{n+1,\nbar}(x,x_1...x_n, \xbar_1...\xbar_\nbar)\\
a^\dagger(f)\psi^{n\nbar}(x_1...x_n, \xbar_1...\xbar_\nbar)
&=\frac{1}{\sqrt{n}}\sum_{j=1}^n(-1)^{j+1} f(x_j) \: \psi^{n-1,\nbar}(x_1...\widehat{x_j}...\xbar_\nbar)\\
b(g)\psi^{n\nbar}(x_1...x_n, \xbar_1...\xbar_\nbar)
&=(-1)^n\sqrt{\nbar+1}\sum_{\sbar=1}^4\int\!\! d^3\vxbar\: Cg(\xbar)^* \: \psi^{n,\nbar+1}(x_1...x_n,\xbar, \xbar_1...\xbar_\nbar)\\
b^\dagger(g)\psi^{n\nbar}(x_1...x_n, \xbar_1...\xbar_\nbar)
&=\frac{(-1)^n}{\sqrt{\nbar}}\sum_{\jbar=1}^\nbar (-1)^{\jbar+1} Cg(\xbar_\jbar) \: \psi^{n-1,\nbar}(x_1...\widehat{\xbar_\jbar}...\xbar_\nbar)
\end{align}
\end{subequations}
Then, for $f\in \Hilbert_1$, the field operator is (e.g., \cite{Schwe61,BD2,Scha})
\be
\Psi(f) = a(P_{1+}f) + b^\dagger(P_{1-}f)\,,
\ee
where $P_{1\pm}$ is the projection $\Hilbert_1\to\Hilbert_{1\pm}$. The field operators satisfy canonical anti-commutation relations (CAR) for $f_1,f_2\in \Hilbert_1$,
\begin{subequations}\label{CAR}
\begin{align}
\{\Psi(f_1),\Psi(f_2)\} &= \{\Psi^\dagger(f_1), \Psi^\dagger(f_2)\}=0,\\
\{\Psi(f_1),\Psi^\dagger(f_2)\} &= \scp{f_1}{f_2}_{\Hilbert_1}~I.
\end{align}
\end{subequations}
The expression $\Psi_s(\vx)$ can then be understood as an operator-valued distribution.

The charge operators $Q(A)$ for $A\subseteq \RRR^3$ are then defined as
\begin{align}
Q(A) 
&:= \int_A d^3\vx \: Q(\vx) = -\int_A d^3\vx \sum_{s=1}^4 : \Psi^\dagger(x) \, \Psi(x): \\
&= -\int_A d^3\vx \sum_{s=1}^4 \Psi^\dagger(x) \, \Psi(x) + c\,,
\end{align}
where the colons mean normal ordering, which amounts here merely to the addition of an infinite constant $c$. (The sign in front of the integral reflects the fact that the charge of the electron is negative.) The charge operators are apparently self-adjoint with spectrum $\ZZZ$, $\sigma$-additive, i.e.,
\be
Q(A_1\cup A_2\cup \ldots)=Q(A_1)+Q(A_2)+\ldots
\text{ if }A_i\cap A_j=\emptyset
\text{ for }i\neq j,
\ee
and commute with each other by virtue of the CAR \eqref{CAR},
\be\label{Qcommute}
[Q(A),Q(A')]=0~~~~\forall A,A' \subseteq \RRR^3.
\ee
By the spectral theorem, they can be diagonalized simultaneously.

\begin{defn}
$P_\nat$ is the PVM on the joint spectrum of all $Q(A)$, $A\subseteq \RRR^3$, that arises from diagonalizing all $Q(A)$ simultaneously.
\end{defn}

A point in the joint spectrum provides an eigenvalue $q(A)$ of $Q(A)$ for every $A\subseteq\RRR^3$, and thus an integer-valued set function $A\mapsto q(A)$ that inherits the additivity from the $Q(A)$ (because for a joint eigenvector of two operators $Q_1,Q_2$, the eigenvalue of the sum $Q_1+Q_2$ is the sum of the eigenvalues).

\begin{conj}\label{conj}
The joint spectrum consists (up to $P_\nat$-null sets) of locally bounded functions $A\mapsto q(A)$.
\end{conj}

This means the following. A set function $A\mapsto q(A)$ is \emph{locally bounded} if for every $\vx\in\RRR^3$ there is a radius $r>0$ and a (finite) $C>0$ such that for all $A\subseteq B_r(\vx)=\{\vy\in\RRR^3:|\vy-\vx|<r\}$,
\be
|q(A)|<C\,. 
\ee

Here is how an integer-valued, additive set function $q(\cdot)$ can fail to be locally bounded: suppose $A_0$ is a brick (3d rectangle) with (say) $q(A_0)=0$ that can be subdivided into two smaller bricks, $A_0=A_1 \cup A'_1$, such that $q(A_1)=1$ and $q(A'_1)=-1$; suppose $A_1$ can be further subdivided into two smaller bricks, $A_1=A_2 \cup A'_2$, such that $q(A_2)=2$ and $q(A'_2)=-1$; and so on, $A_k=A_{k+1} \cup A'_{k+1}$ such that $q(A_k)=k$ and $q(A'_k)=-1$. Obviously, the $q(A_k)$ are unbounded while all $A_k$ are contained in $A_0$ and thus in some ball. 

It also follows that the $A'_k$ are mutually disjoint, so $q(A'_1\cup A'_2\cup \ldots)$, which must be a (finite) integer, cannot agree with $q(A'_1) +q(A'_2)+\ldots = -\infty$, so $q(\cdot)$ cannot be $\sigma$-additive, while it can be finitely additive. In fact, Conjecture~\ref{conj} can equivalently be reformulated by saying that \emph{the joint spectrum consists (up to $P_\nat$-null sets) of locally $\sigma$-additive functions $q(A)$.}\footnote{It may seem strange that $\sigma$-additivity would locally be inherited but not globally. Why would that be? It may be because wave functions tend to 0 at infinity.}

Now locally bounded, additive set functions are of the form
\be
q(A) = \sum_{\vx \in A\cap S} q_{\vx}
\ee
for every bounded set $A$, where $S$ is a locally finite subset of $\RRR^3$ and $q_{\vx}$ an integer (the charge at $\vx$). In other words, if the charge content of $A$ is locally bounded, then there are only locally finitely many charges. Since it is an event of measure zero for two charges to be at the same location, we can assume as well that $q_{\vx}= \pm 1$. That is, for a point in the joint spectrum given by the set function $q(\cdot)$, there is a locally finite set $S_+$ of positive charges and a locally finite set $S_-$ of negative charges so that
\be
q(A) = \#(A\cap S_+) - \#(A\cap S_-)
\ee
for every bounded set $A\subseteq \RRR^3$. So, a configuration can be specified by specifying $S_-$ and $S_+$; correspondingly, $P_\nat$ is a PVM on 
\be
\Q_\lf = \Gamma_\lf(\RRR^3) \times \Gamma_\lf(\RRR^3)
\ee
with
\be
\Gamma_\lf(\RRR^3)=\Bigl\{q\subset \RRR^3: \#(q\cap B_r(\vx))<\infty \: \forall r>0 \: \forall \vx \in \RRR^3\Bigr\}
\ee
the space of locally finite configurations.

\section{Definition of $P_\nat$ in Finite Volume}
\label{sec:Pnatfinite}

Plausibly, our universe is approximately a 3-sphere and thus has finite 3-volume. Finite 3-volume makes $P_\nat$ easier to treat, as then configurations are finite. 

\begin{figure}[h]
\begin{center} 
  \begin{tikzpicture}
  \draw[color=gray] (0,0) -- (0,4) -- (4,4) -- (4,0) -- cycle;
  \filldraw[color=blue] (2.6278,3.3109) circle [radius=0.04];
  \filldraw[color=blue] (2.4832,1.1886) circle [radius=0.04];
  \filldraw[color=red] (0.3327,1.8012) circle [radius=0.04];
  \filldraw[color=red] (3.1720,0.4185) circle [radius=0.04];
  \end{tikzpicture}
\end{center}
 \caption{Example of a configuration in $\Gamma(\TTT^2)\times \Gamma(\TTT^2)$; electrons are marked in blue, positrons in red.}\label{fig:ex}
\end{figure}
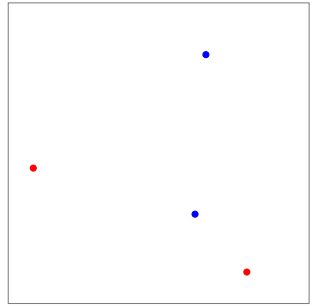

As a concrete model with finite 3-volume, let us consider instead of $\RRR^3$ a box $[0,L)^3$ of side length $L$, and
\be
H_1=-i\valpha\cdot \nabla+\beta m
\ee
with periodic boundary conditions, so that space can be regarded as a 3-torus $\TTT^3$ of period length $L$. Define $\Hilbert_1=L^2(\TTT^3,\CCC^4)$, $\Hilbert_{1\pm}$ as the positive (negative) spectral subspace of $H_1$, $\Hilbert=\Fock(\Hilbert_{1+})\otimes \Fock(C\Hilbert_{1-})$, $\Psi_s(\vx)$, $Q(A)$ for $A\subseteq \TTT^3$, and $P_\nat$ as before. Conjecture~\ref{conj} then implies that $P_\nat$ is concentrated on the finite configurations (see Figure~\ref{fig:ex}), i.e., it is a PVM on
\be\label{confTTT3}
\conf=\Gamma(\TTT^3)\times \Gamma(\TTT^3)\,.
\ee

I propose an ontology, corresponding to this configuration space, with a variable but finite number of point particles that come in two kinds called electrons and positrons. The Born rule then says that their configuration is distributed according to
\be
\PPP_\nat(B) = \scp{\psi}{P_\nat(B)|\psi}~~~\forall B\subseteq \Gamma(\TTT^3)\times\Gamma(\TTT^3).
\ee
I will discuss the possibility of Bohmian trajectories for them in Section~\ref{sec:Bohm}. We will see that $P_\nat$ is unusual in that its vacuum subspace, the range of $P_\nat(\{\emptyset\})$, has infinite dimension rather than dimension 1.

\section{Alternative Approach to $P_\nat$ via Discretization}
\label{sec:discrete}

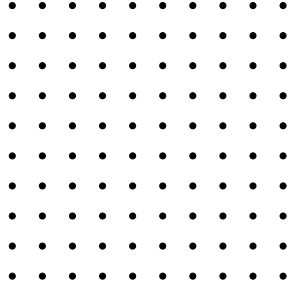
\begin{figure}[h]
\begin{center} 
  \begin{tikzpicture}
  \foreach \x in {1,...,10}
     \foreach \y in {1,...,10}
     {
       \filldraw (\x*0.4,\y*0.4) circle [radius=0.04];
     }
  \end{tikzpicture}
\end{center}
 \caption{Sites of the lattice $\lattice$ in 2 dimensions for $N=10$}\label{fig:lattice}
\end{figure}

It is useful to see how $P_\nat$ arises in the continuum limit of a discrete space. To this end, consider, instead of $\RRR^3$ or $\TTT^3$, a discrete version of $\TTT^3$ in the form of the finite lattice (see Figure~\ref{fig:lattice})
\be
\lattice := \bigl([0,L) \cap \tfrac LN \ZZZ\bigr)^3
\ee
with $N^3$ sites.

Let the 1-particle Hamiltonian be
\be
H_1=-i\valpha\cdot \nabla+\beta m
\ee
with $\nabla$ the difference operator,
\be
\nabla_j \psi(\vx) = \frac{\psi(\vx+\tfrac LN \ve_j)-\psi(\vx)}{\tfrac LN},
\ee
and periodic boundaries. The 1-particle Hilbert space $\Hilbert_1 = L^2(\lattice,\CCC^4)$ has dimension $4N^3$ while the Fock space $\Hilbert= \Fock(\Hilbert_1)$ has dimension $16^{N^3}$. The latter can be regarded as the tensor product
\be
\Hilbert = \bigotimes_{\vx\in\lattice} \Hilbert_x\,,
\ee
where
\be
\Hilbert_x = \Fock(\CCC^4)\,.
\ee
Decomposing the latter Fock space into sectors by particle number yields
\begin{align}
\Hilbert_x
&=\Fock_0\oplus \Fock_1 \oplus \Fock_2 \oplus \Fock_3 \oplus \Fock_4\\
\text{with dimension } 16 
&= \,~1~+~4~\,+~6~\,+~4~\,+~1\,.
\end{align}
In this setting, $\Psi_s(\vx)$ is literally the annihilation operator and $\Psi^\dagger_s(\vx)$ literally the creation operator.

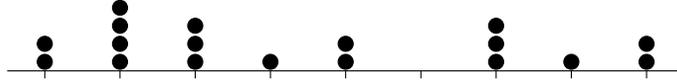
\begin{figure}[h]
\begin{center} 
  \begin{tikzpicture}
  \draw (-0.5,0)--(8.5,0);
  \draw (0,-0.1)--(0,0);
  \draw (1,-0.1)--(1,0);
  \draw (2,-0.1)--(2,0);
  \draw (3,-0.1)--(3,0);
  \draw (4,-0.1)--(4,0);
  \draw (5,-0.1)--(5,0);
  \draw (6,-0.1)--(6,0);
  \draw (7,-0.1)--(7,0);
  \draw (8,-0.1)--(8,0);
  \filldraw[color=black] (0,0.12) circle [radius=0.1];
  \filldraw[color=black] (0,0.36) circle [radius=0.1];
  \filldraw[color=black] (1,0.12) circle [radius=0.1];
  \filldraw[color=black] (1,0.36) circle [radius=0.1];
  \filldraw[color=black] (1,0.6) circle [radius=0.1];
  \filldraw[color=black] (1,0.84) circle [radius=0.1];
  \filldraw[color=black] (2,0.12) circle [radius=0.1];
  \filldraw[color=black] (2,0.36) circle [radius=0.1];
  \filldraw[color=black] (2,0.6) circle [radius=0.1];
  \filldraw[color=black] (3,0.12) circle [radius=0.1];
  \filldraw[color=black] (4,0.12) circle [radius=0.1];
  \filldraw[color=black] (4,0.36) circle [radius=0.1];
  \filldraw[color=black] (6,0.12) circle [radius=0.1];
  \filldraw[color=black] (6,0.36) circle [radius=0.1];
  \filldraw[color=black] (6,0.6) circle [radius=0.1];
  \filldraw[color=black] (7,0.12) circle [radius=0.1];
  \filldraw[color=black] (8,0.12) circle [radius=0.1];
  \filldraw[color=black] (8,0.36) circle [radius=0.1];
  \end{tikzpicture}
\end{center}
 \caption{Example of a configuration on the 1d lattice. Lattice sites are marked by ticks on the space axis; each site can be occupied by up to 4 particles.}\label{fig:pre1}
\end{figure}

An element $\psi$ of $\Fock(\Hilbert_1)$ can be written as a function of a variable number $\ell$ of particles, $\psi = \psi^{\ell}_{s_1...s_\ell}(\vx_1\ldots \vx_\ell)$, so $|\psi|^2$ defines a probability distribution over the configuration space
\be
\conf_N=\{0,1,2,3,4\}^\lattice\,,
\ee
as unordered configurations can be specified by specifying the occupation number for each lattice site, and the maximum occupation number is 4 due to fermionic symmetry and the dimension 4 of the spin space $\CCC^4$; see Figure~\ref{fig:pre1} for an example of a configuration from $\conf_N$.

\begin{figure}[h]
\begin{center} 
  \begin{tikzpicture}
  \draw (-0.5,0)--(8.5,0);
  \draw (0,-0.1)--(0,0);
  \draw (1,-0.1)--(1,0);
  \draw (2,-0.1)--(2,0);
  \draw (3,-0.1)--(3,0);
  \draw (4,-0.1)--(4,0);
  \draw (5,-0.1)--(5,0);
  \draw (6,-0.1)--(6,0);
  \draw (7,-0.1)--(7,0);
  \draw (8,-0.1)--(8,0);
  \filldraw[color=black] (0,0.12) circle [radius=0.1];
  \filldraw[color=black] (0,0.36) circle [radius=0.1];
  \filldraw[color=black] (1,0.12) circle [radius=0.1];
  \filldraw[color=black] (1,0.36) circle [radius=0.1];
  \filldraw[color=black] (2,0.12) circle [radius=0.1];
  \filldraw[color=black] (2,0.36) circle [radius=0.1];
  \filldraw[color=black] (3,0.12) circle [radius=0.1];
  \filldraw[color=black] (3,0.36) circle [radius=0.1];
  \filldraw[color=black] (4,0.12) circle [radius=0.1];
  \filldraw[color=black] (4,0.36) circle [radius=0.1];
  \filldraw[color=black] (5,0.12) circle [radius=0.1];
  \filldraw[color=black] (5,0.36) circle [radius=0.1];
  \filldraw[color=black] (6,0.12) circle [radius=0.1];
  \filldraw[color=black] (6,0.36) circle [radius=0.1];
  \filldraw[color=black] (7,0.12) circle [radius=0.1];
  \filldraw[color=black] (7,0.36) circle [radius=0.1];
  \filldraw[color=black] (8,0.12) circle [radius=0.1];
  \filldraw[color=black] (8,0.36) circle [radius=0.1];
  \end{tikzpicture}
\end{center}
 \caption{The ``level configuration'' $q_\sL$ on the 1d lattice}\label{fig:pre2}
\end{figure}
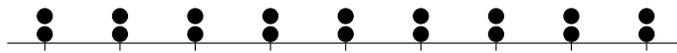

Let me introduce some terminology. I call the ``particles'' in this configuration \emph{pre-particles} because in my proposal they are not ontologically the real particles. I call the associated PVM $P_\pre$; it is a PVM on $\conf_N$ acting on $\Hilbert$; in fact, $\Hilbert$ can be regarded as a subspace of $\Kilbert = L^2(\conf_N, \sC)$ with $\sC=\oplus_{n=0}^{4N^3}(\CCC^4)^{\otimes n}$, and
\be
P_\pre(B) = P_\Hilbert 1_B P_\Hilbert\,.
\ee
The \emph{bottom configuration} is the unique configuration with 0 pre-particles. The \emph{bottom state} $|B\rangle$ is the unique (up to phase) unit vector in $\Hilbert$ that has 0 pre-particles. The \emph{level configuration} $q_\sL$ is the unique configuration with 2 pre-particles at each lattice site (see Figure~\ref{fig:pre2}).

\begin{figure}[h]
\begin{center} 
  \begin{tikzpicture}[scale=0.6]
  \filldraw[color=gray] (-0.4,1)--(0,1)--(0,2)--(-0.4,2)--cycle;
  \filldraw[color=gray] (-0.4,3)--(0,3)--(0,4)--(-0.4,4)--cycle;
  \filldraw[color=gray] (0,0)--(1,0)--(1,1)--(0,1)--cycle;
  \filldraw[color=gray] (0,2)--(1,2)--(1,3)--(0,3)--cycle;
  \filldraw[color=gray] (1,0)--(2,0)--(2,1)--(1,1)--cycle;
  \filldraw[color=gray] (1,1)--(2,1)--(2,2)--(1,2)--cycle;
  \filldraw[color=gray] (2,0)--(3,0)--(3,1)--(2,1)--cycle;
  \filldraw[color=gray] (2,2)--(3,2)--(3,3)--(2,3)--cycle;
  \filldraw[color=gray] (3,0)--(4,0)--(4,1)--(3,1)--cycle;
  \filldraw[color=gray] (3,3)--(4,3)--(4,4)--(3,4)--cycle;
  \filldraw[color=gray] (4,1)--(5,1)--(5,2)--(4,2)--cycle;
  \filldraw[color=gray] (4,2)--(5,2)--(5,3)--(4,3)--cycle;
  \filldraw[color=gray] (5,0)--(6,0)--(6,1)--(5,1)--cycle;
  \filldraw[color=gray] (5,2)--(6,2)--(6,3)--(5,3)--cycle;
  \filldraw[color=gray] (6,1)--(7,1)--(7,2)--(6,2)--cycle;
  \filldraw[color=gray] (6,2)--(7,2)--(7,3)--(6,3)--cycle;
  \filldraw[color=gray] (7,0)--(8,0)--(8,1)--(7,1)--cycle;
  \filldraw[color=gray] (7,1)--(8,1)--(8,2)--(7,2)--cycle;
  \filldraw[color=gray] (8,1)--(8.4,1)--(8.4,2)--(8,2)--cycle;
  \filldraw[color=gray] (8,3)--(8.4,3)--(8.4,4)--(8,4)--cycle;
  \draw (-0.4,0)--(8.4,0);
  \draw (-0.4,1)--(8.4,1);
  \draw (-0.4,2)--(8.4,2);
  \draw (-0.4,3)--(8.4,3);
  \draw (-0.4,4)--(8.4,4);
  \draw (0,0)--(0,4);
  \draw (1,0)--(1,4);
  \draw (2,0)--(2,4);
  \draw (3,0)--(3,4);
  \draw (4,0)--(4,4);
  \draw (5,0)--(5,4);
  \draw (6,0)--(6,4);
  \draw (7,0)--(7,4);
  \draw (8,0)--(8,4);
  \draw (0.5,-0.1)--(0.5,0);
  \draw (1.5,-0.1)--(1.5,0);
  \draw (2.5,-0.1)--(2.5,0);
  \draw (3.5,-0.1)--(3.5,0);
  \draw (4.5,-0.1)--(4.5,0);
  \draw (5.5,-0.1)--(5.5,0);
  \draw (6.5,-0.1)--(6.5,0);
  \draw (7.5,-0.1)--(7.5,0);
  \end{tikzpicture}
\end{center}
 \caption{Schematic representation of a choice of two basis vectors from $\{e_1,e_2,e_3,e_4\}$ for each site $\vx$ in a 1d lattice. Ticks mark lattice sites on the space axis (to the right), while the elements of $\{e_1,e_2,e_3,e_4\}$ are symbolized by boxes ordered upward.}\label{fig:pre3}
\end{figure}
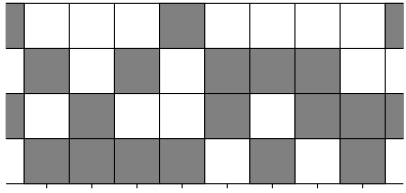

A \emph{level state} is a vector in $\Hilbert$ concentrated on the level configuration. \emph{Level space} $\sL$ is the set of all level states; it is a subspace of $\Hilbert$ given by
\be
\sL= \bigotimes_{\vx \in \lattice} \Fock_2(\CCC^4)= \text{range }P_\pre(\{q_\sL\})\,.
\ee 
Its dimension is $6^{N^3}$. Here is what $\sL$ looks like:
Let $\{e_1,e_2,e_3,e_4\}$ denote the canonical orthonormal basis of $\CCC^4$ and $\ket{ij}:=e_i\wedge e_j$. 
Then $\{\ket{12},\ket{13},\ket{14},\ket{23},\ket{24},\ket{34}\}$ is an orthonormal basis of $\Fock_2(\CCC^4)$, and
\be
\Bigl\{\bigotimes_{\vx\in\lattice}\ket{i_{\vx} j_{\vx}}:1\leq i_{\vx}<j_{\vx}\leq 4~~\forall \vx\in\lattice\Bigr\}
\ee 
is an orthonormal basis of $\sL$; an example of a choice of $i_{\vx}$ and $j_{\vx}$ for every $\vx\in\lattice$ is shown in Figure~\ref{fig:pre3}.

The sea state is $\ket{\Omega} = \varphi_1 \wedge\varphi_2 \wedge \cdots \wedge \varphi_{2N^3}$ (up to phase) with $\{\varphi_1,\varphi_2,\ldots,\varphi_{2N^3}\}$ any orthonormal basis of the negative spectral subspace $\Hilbert_{1-}$ of $H_1$.
The sea state is not level, $\ket{\Omega} \notin \sL$. It is an eigenstate of $\Psi^\dagger(\varphi_k)\Psi(\varphi_k)$, but not of $\Psi_s^\dagger(\vx) \Psi_s(\vx)$.

\begin{figure}[h]
\begin{center} 
  \begin{tikzpicture}
  \node at (-1,.5) {(a)};
  \draw (-0.5,0)--(8.5,0);
  \draw (0,-0.1)--(0,0);
  \draw (1,-0.1)--(1,0);
  \draw (2,-0.1)--(2,0);
  \draw (3,-0.1)--(3,0);
  \draw (4,-0.1)--(4,0);
  \draw (5,-0.1)--(5,0);
  \draw (6,-0.1)--(6,0);
  \draw (7,-0.1)--(7,0);
  \draw (8,-0.1)--(8,0);
  \filldraw[color=black] (0,0.12) circle [radius=0.1];
  \filldraw[color=black] (0,0.36) circle [radius=0.1];
  \filldraw[color=black] (1,0.12) circle [radius=0.1];
  \filldraw[color=black] (1,0.36) circle [radius=0.1];
  \filldraw[color=black] (1,0.6) circle [radius=0.1];
  \filldraw[color=black] (1,0.84) circle [radius=0.1];
  \filldraw[color=black] (2,0.12) circle [radius=0.1];
  \filldraw[color=black] (2,0.36) circle [radius=0.1];
  \filldraw[color=black] (2,0.6) circle [radius=0.1];
  \filldraw[color=black] (3,0.12) circle [radius=0.1];
  \filldraw[color=black] (4,0.12) circle [radius=0.1];
  \filldraw[color=black] (4,0.36) circle [radius=0.1];
  \filldraw[color=black] (6,0.12) circle [radius=0.1];
  \filldraw[color=black] (6,0.36) circle [radius=0.1];
  \filldraw[color=black] (6,0.6) circle [radius=0.1];
  \filldraw[color=black] (7,0.12) circle [radius=0.1];
  \filldraw[color=black] (8,0.12) circle [radius=0.1];
  \filldraw[color=black] (8,0.36) circle [radius=0.1];
  \end{tikzpicture}

  \bigskip

  \begin{tikzpicture}
  \node at (-1,.5) {(b)};
  \draw (-0.5,0)--(8.5,0);
  \draw (0,-0.1)--(0,0);
  \draw (1,-0.1)--(1,0);
  \draw (2,-0.1)--(2,0);
  \draw (3,-0.1)--(3,0);
  \draw (4,-0.1)--(4,0);
  \draw (5,-0.1)--(5,0);
  \draw (6,-0.1)--(6,0);
  \draw (7,-0.1)--(7,0);
  \draw (8,-0.1)--(8,0);
  \filldraw[color=blue] (1,0.12) circle [radius=0.1];
  \filldraw[color=blue] (1,0.36) circle [radius=0.1];
  \filldraw[color=blue] (2,0.12) circle [radius=0.1];
  \filldraw[color=red] (3,0.12) circle [radius=0.1];
  \filldraw[color=red] (5,0.12) circle [radius=0.1];
  \filldraw[color=red] (5,0.36) circle [radius=0.1];
  \filldraw[color=blue] (6,0.12) circle [radius=0.1];
  \filldraw[color=red] (7,0.12) circle [radius=0.1];
  \end{tikzpicture}
\end{center}
 \caption{The ontology associated with the pre-particle configuration (a) is shown in (b).}\label{fig:pre4}
\end{figure}
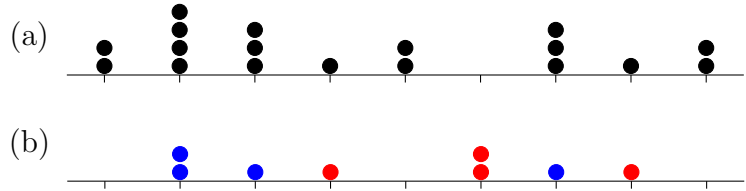

I can now define the analog $P_{\nat,N}$ of $P_\nat$ for the lattice case. On a discrete space (such as $\conf_N$), a PVM can be specified by specifying it for each point. $P_{\nat,N}$ arises from $P_\pre$ by a re-interpretation as in Figure~\ref{fig:pre4}: 2 pre-particles are regarded as 0 electrons and 0 positrons, 3 pre-particles as 1 electron and 0 positrons, 4 pre-particles as 2 electrons and 0 positrons, 1 pre-particle as 0 electrons and 1 positron, and 0 pre-particles as 0 electrons and 2 positrons.

\begin{defn} We represent a configuration by the charge $q(\vx)$ at each $\vx\in\lattice$ and define
\be\label{PnatNdef}
P_{\nat,N}(q) := P_\pre(q_\sL-q)
\ee
for all $q\in \{-2,-1,0,1,2\}^\lattice$.
\end{defn}

Note that range $P_{\nat,N}(\{\emptyset\})=\sL$, so the ``vacuum subspace'' according to $P_\nat$ has large dimension. We obtain $P_\nat$ from $P_{\nat,N}$ in the continuum limit $N\to \infty$ while keeping $L$ fixed, so that $\lattice$ approaches $\TTT^3$, while considering only states that differ from $|\Omega\rangle$ by finitely many pre-particles. Conjecture~\ref{conj} then implies that
\be
\scp{\Omega}{P_{\nat,N}(\{\emptyset\})|\Omega} \xrightarrow{N\to\infty} \scp{\Omega}{P_\nat(\{\emptyset\})|\Omega}\in (0,1)
\ee
while
\be
\scp{\Omega}{P_\nat(\{\emptyset\})|\Omega}\xrightarrow{L\to\infty}0.
\ee

\section{Dirac Sea}
\label{sec:Diracsea}

\begin{figure}[h]
\begin{center}
\includegraphics[width=4cm]{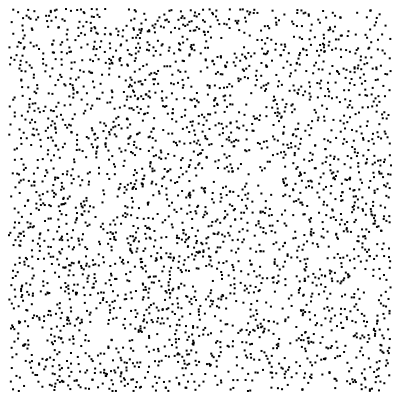}
\end{center}
\caption{A random set of points}
\label{fig:points1}
\end{figure}

Several authors \cite{BH,CS07,DEO17} have proposed, specifically in the context of Bohmian mechanics, that the Dirac sea should be taken literally in the sense that positrons do not exist as particles in their own right, whereas electrons of positive or negative energy do; a situation that we normally regard as vacuum would then not only have a quantum state $|\Omega\rangle$ that in some mathematical sense involves an infinitude of particles but also have in 3-dimensional reality an actual infinitude of Bohmian particles. While it is perhaps not fully clear what such a configuration should look like, I always took it to be a countable dense set in $\RRR^3$, like the rational points, and imagined it to look like Figure~\ref{fig:points1}, just denser.

Now the way I have drawn Figure~\ref{fig:points1}, it is a realization of a Poisson point process, i.e., a large number of independent uniformly distributed random points, whereas the joint distribution of many electrons with a wave function given by the wedge product of negative-energy eigenstates is not uniform in configuration space, as it vanishes at configurations with more than 4 particles at the same location; this fact should suppress clusters (i.e., many nearby particles), lead to an effective repulsion between the particles, and favor more evenly distributed configurations. So the correct picture should look more uniform than Figure~\ref{fig:points1}.

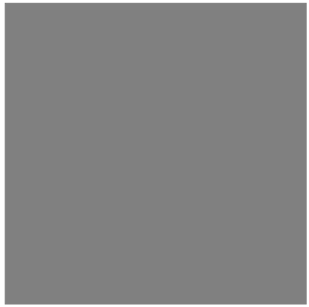
\begin{figure}[h]
\begin{center} 
  \begin{tikzpicture}
  \filldraw[color=gray] (0,0) -- (0,4) -- (4,4) -- (4,0) -- cycle;
  \end{tikzpicture}
\end{center}
 \caption{In the continuum limit, the level configuration (i.e., occupation number 2 at every location) would correspond to a uniform density of matter.}\label{fig:points4}
\end{figure}

Now the most uniform configuration conceivable would have exactly 2 particles at each location, as depicted in Figure~\ref{fig:pre2} for a lattice. In the continuum limit, the appropriate figure for that, to compare with Figure~\ref{fig:points1}, would seem to be a uniform shade of grey over all space, as depicted in Figure~\ref{fig:points4}. 

\begin{figure}[h]
\begin{center} 
  \begin{tikzpicture}
  \filldraw[color=gray] (0,0) -- (0,4) -- (4,4) -- (4,0) -- cycle;
  \filldraw[color=black] (2.6278,3.3109) circle [radius=0.04];
  \filldraw[color=black] (2.4832,1.1886) circle [radius=0.04];
  \filldraw[color=white] (0.3327,1.8012) circle [radius=0.04];
  \filldraw[color=white] (3.1720,0.4185) circle [radius=0.04];
  \end{tikzpicture}
\end{center}
 \caption{A typical configuration of the sea state on a lattice with invisibly small width according to Conjecture~\ref{conj}; occupation number 2 of a lattice point is represented in grey, occupation number $\geq 3$ in black, occupation number $\leq 1$ in white.}\label{fig:points2}
\end{figure}
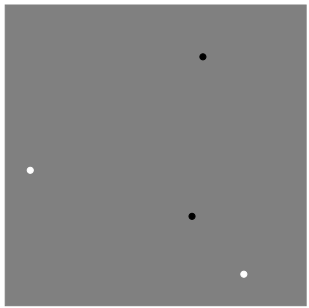

Thus, the correct picture should lie between Figures~\ref{fig:points1} and \ref{fig:points4}. If Conjecture~\ref{conj} is correct, then we can say what the correct picture looks like: not at all like Figure~\ref{fig:points1} but like Figure~\ref{fig:points2}.

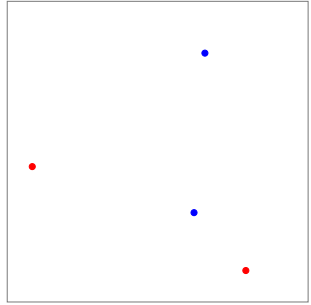
\begin{figure}[h]
\begin{center} 
  \begin{tikzpicture}
  \draw[color=gray] (0,0) -- (0,4) -- (4,4) -- (4,0) -- cycle;
  \filldraw[color=blue] (2.6278,3.3109) circle [radius=0.04];
  \filldraw[color=blue] (2.4832,1.1886) circle [radius=0.04];
  \filldraw[color=red] (0.3327,1.8012) circle [radius=0.04];
  \filldraw[color=red] (3.1720,0.4185) circle [radius=0.04];
  \end{tikzpicture}
\end{center}
 \caption{A typical configuration of the sea state according to $P_\nat$; electrons are marked in blue, positrons in red.}\label{fig:points3}
\end{figure}

That is, for a lattice with large $N$, the great majority of sites is occupied by 2 particles while only a number of sites much smaller than $N$ (one that remains finite in the limit $N\to\infty$) has occupation number different from 2. But in view of Figure~\ref{fig:points2}, it seems natural to regard points with occupation number $\geq 3$ as electrons, those with $\leq 1$ as positrons, and those with $2$ as vacuum, i.e., to replace Figure~\ref{fig:points2} by Figure~\ref{fig:points3}.

In sum, the ``sea view'' (that the Dirac sea should be taken literally) gets pushed by Conjecture~\ref{conj} towards the ontology associated with $P_\nat$.


It is striking that the ontology suggested by Figure~\ref{fig:points1} breaks the symmetry between electrons and positrons: electrons are real and positrons are not. This symmetry is restored in the ontology associated with $P_\nat$. In fact, as Bjorken and Drell wrote \cite[Sec.~5.2]{BD1}, we could have started equally well with the Dirac equation for the positrons (which amounts to using $\overline{\Hilbert_1}$ instead of $\Hilbert_1$) and would have arrived at the same theory. Here, of course, we go further than Bjorken and Drell and find the symmetry still respected in the position operators and, as we will see below, in the Bohmian trajectories.

\section{Difficulty with $P_\obv$}
\label{sec:Pobv}

In this section, I present reasons for thinking that with $P_\obv$, \emph{interaction locality} (i.e., the absence of interaction terms between spacelike separated regions in the unitary time evolution) will not hold strictly but at most approximately. Some authors (e.g., Beck \cite[Sec.~5.7]{Beck}) have argued that interaction locality should not be expected to hold strictly. Likewise with \emph{propagation locality} (i.e., the absence of faster-than-light propagation of wave functions):\footnote{The Newton-Wigner position observable \cite{NW49}, which will not be considered here, is known \cite{Tha} to violate propagation locality already in the 1-particle case.} some authors (e.g., Soffer \cite{Sof11}) do not expect it to hold strictly. I tend to think that both hold strictly in our universe.

Be that as it may, here are two reasons for expecting $P_\obv$ to lead to conflicts with interaction locality.

The first reason concerns the question how a detector, if $P_\obv$ yields the detection probabilities, should collapse the wave function. Specifically, a function $f\in\Hilbert_{1+}$ that vanishes in an open set vanishes everywhere \cite[Corollary 1.7]{Tha}, so $\rho_\obv$ can never have compact support. So consider a quantum state $\psi$ concentrated in the 1-electron, 0-positron sector according to $P_\obv$, suppose we apply a detector at $t=0$ and detect the particle in a compact region $A\subset \RRR^3$, say, a closed ball. Upon detection, the wave function should collapse. While it is not clear what exactly the post-collapse wave function $\psi'$ should be, it is clear that the $(1,0)$-sector of $\psi'$ will not be concentrated in $A$ but will have tails outside. In particular, there is positive probability of finding the particle far from $A$ at any $t>0$. Since with $P_\obv$ wave functions do not propagate faster than light (in fact, the Bohmian particle does not move faster than light), this leads to the conclusion that with positive probability, the particle actually was not in $A$ at $t=0$, so the detector in $A$ was triggered by a particle spacelike separated from $A$ in violation of interaction locality.

Here is the second reason: it seems that no conceivable interaction term would obey interaction locality with $P_\obv$ while conserving particle number, as illustrated by the following example. Assuming otherwise, consider a 2-electron wave function and let us write $\psi$ for the $(2,0)$-sector, $\psi=\psi^{20}_{s_1s_2}(\vx_1,\vx_2)$; consider $\psi$ at times $0$ and $t$. Let $F_t=\exp(-iH_0t)$ be the free time evolution. By interaction locality, $\psi_t$ agrees with $F_t\psi_0$ outside the $t$-neighborhood of the diagonal $\{(\vx_1,\vx_2): \vx_1=\vx_2\}$ in configuration space $\RRR^6$. If both $\psi_t$ and $\psi_0$ lie in $\Hilbert_{1+}\otimes \Hilbert_{1+}$, then so does $\psi_t-F_t\psi_0$. But the only function in $\Hilbert_{1+}\otimes \Hilbert_{1+}$ that vanishes outside the $t$-neighborhood of the diagonal is 0. So, the evolution agrees with $F_t$, in contradiction to the assumption of interaction.

I close this section pointing out the widespread use of $P_\obv$. While it is standard in QFT to never mention a POVM, often remarks about how to interpret $\psi\in\Hilbert$ are inspired by $P_\obv$:

\begin{itemize}
\item Thaller \cite{Tha} p.~277: ``the probability that there are just $n$ particles and [$\nbar$] antiparticles at a given time is [$\|\psi^{n\nbar}\|^2$]''

\item Thaller \cite{Tha} p.~277: ``[$\ket{\Omega}$] describes the possibility that there are no particles at all''

\item Schweber \cite{Schwe61} p.~230: ``The basis vectors [$a^\dagger(P_{1+}x_1)\cdots a^\dagger(P_{1+}x_n)b^\dagger(P_{1-}\xbar_1) \cdots$ $b^\dagger(P_{1-}\xbar_\nbar)\ket{\Omega}$] span the states in which there exist [$n$] particles and [$\nbar$] antiparticles.''

\item Schweber \cite{Schwe61} p.~231: ``[$\psi^{01}(x)$] is the probability amplitude for finding the antiparticle [at $\vx$].''
\end{itemize}

None of these statements would be correct about $P_\nat$, all of them are correct about $P_\obv$. 

I am not quoting these authors to criticize them. They made the statements I am quoting because they wanted to be clear and explicit about the physical meaning of certain quantum states; and for this they used $P_\obv$ or equivalent concepts, which is why I am quoting them.

\section{Properties of $P_\nat$}
\label{sec:properties}

\subsection{Relations with $Q(A)$ and $\Psi_s(\vx)$}

We collect several facts; for convenience, we take 3-space to be $\TTT^3$. 
To begin with, as with every PVM, 
\be
[P_\nat(B),P_\nat(B')]=0
\ee
for all $B,B'\subseteq \conf=\Gamma(\TTT^3)\times \Gamma(\TTT^3)$.

Second, on $\conf$ one can define, for any $A\subseteq \TTT^3$, the \emph{electron number function} in $A$, $n_{\el,A}:\conf \to \NNN\cup\{0\}$,
\be
n_{\el,A}(q_\el,q_\pos) = \# (A\cap q_\el)\,,
\ee
as well as the \emph{positron number function} $n_{\pos,A}:\conf\to\NNN\cup\{0\}$,
\be
n_{\pos,A}(q_\el,q_\pos) = \# (A\cap q_\pos)\,,
\ee
and the \emph{charge content function} $q_A:\conf\to\ZZZ$,
\be
q_A(q_\el,q_\pos)= n_{\pos,A} - n_{\el,A}\,.
\ee
For fixed configuration $(q_\el,q_\pos)$, each of the three depends on $A$ in a $\sigma$-additive way. 

From $P_\nat$, one can then define an \emph{electron number operator} and a \emph{positron number operator} according to
\begin{subequations}
\begin{align}
N_{\nat,\el}(A) &= \int_\conf P_\nat(dq) \: n_{\el,A}(q)\\
N_{\nat,\pos}(A) &= \int_\conf P_\nat(dq) \: n_{\pos,A}(q)\,.
\end{align}
\end{subequations}
We can then express the charge operators as
\begin{subequations}
\begin{align}
Q(A) &= N_{\nat,\pos}(A) - N_{\nat,\el}(A)\\
&= \int_{\conf} P_\nat(dq) \: q_A(q)\,.
\end{align}
\end{subequations}
It is worth noting that for number operators defined in the analogous way from $P_\obv$, we have that
\be\label{NobvQT3}
N_{\obv,\pos}(\TTT^3) - N_{\obv,\el}(\TTT^3) = Q(\TTT^3) = N_{\nat,\pos}(\TTT^3) - N_{\nat,\el}(\TTT^3)\,,
\ee
but not so for $A\neq \TTT^3$.

Next, the CAR \eqref{CAR} imply that
\begin{subequations}
\begin{align}
[Q(A), \Psi_s(\vx)] &=- 1_A(\vx) \, \Psi_s(\vx)\\
[Q(A), \Psi_s^\dagger(\vx)] &= ~~\, 1_A(\vx) \, \Psi_s^\dagger(\vx)\,. 
\end{align}
\end{subequations}
As a consequence, if $\phi$ is an eigenvector of $Q(A)$ with eigenvalue $q$, then $\Psi_s(\vx)\phi$ and $\Psi_s^\dagger(\vx)\phi$ are also eigenvectors of $Q(A)$ with eigenvalue $q\mp 1_A(\vx)$.

The last fact has the following consequence for $P_\nat$. On $\conf=\Gamma(\TTT^3)\times \Gamma(\TTT^3)$, define the \emph{charge lowering mapping} $\ell_{\vx}:\conf\to\conf$ for $\vx\in\TTT^3$ by
\be
\ell_{\vx}(q_\el,q_\pos) = \begin{cases}
(q_\el,q_\pos \setminus \{\vx\})& \text{if }\vx \in q_\pos\\
(q_\el\cup\{\vx\},q_\pos) & \text{if }\vx\notin q_\pos \end{cases}
\ee
and the \emph{charge raising mapping} $r_{\vx}:\conf\to\conf$ by
\be
r_{\vx}(q_\el,q_\pos) = \begin{cases} (q_\el\setminus\{\vx\},q_\pos) & \text{if }\vx\in q_\el\\\
(q_\el,q_\pos\cup\{\vx\}) & \text{if }\vx\notin q_\el\,. \end{cases}
\ee
Then, for every $B\subseteq \conf$ and every $\vx\in\TTT^3$,
\begin{subequations}
\begin{align}
&\Psi_s(\vx) \text{ maps range } P_\nat(B) 
\text{ to range }P_\nat(r_{\vx}(B))\,,\\
&\Psi_s^\dagger(\vx) \text{ maps range } P_\nat(B) 
\text{ to range }P_\nat(\ell_{\vx}(B))\,.
\end{align}
\end{subequations}
(This statement is not rigorously true since $\Psi_s(\vx)$ is not rigorously defined as an operator.)

In the same direction, one can obtain the following explicit formula for $\rho_\nat$ on $\TTT^3$:
\begin{align}
&\rho^{n\nbar}_\nat(\vx_1\ldots \vx_n,\vxbar_1\ldots \vxbar_\nbar) =\nonumber\\
&~~~~~~~~~~~~~~~~~~~\frac{1}{3^{n+\nbar}}\sum_{\substack{s_1...s_n\\\sbar_1...\sbar_\nbar}}
\scp{\psi}{\Psi^\dagger(x_1)\cdots\Psi^\dagger(x_n)\Psi(\xbar_1)\cdots \Psi(\xbar_\nbar) ~\times\nonumber\\
&~~~~~~~~~~~~~~~~~~~\times~~P_\nat(\{\emptyset\})\Psi^\dagger(\xbar_\nbar)\cdots \Psi^\dagger(\xbar_1)\Psi(x_n)\cdots \Psi(x_1)|\psi},
\end{align}
provided the locations are pairwise distinct.

A last relation concerns the \emph{algebra of local observables} or \emph{local algebra} \cite{Haa92}. Usually, this algebra is considered for an open subset $\open$ of \emph{space-time}; here, instead, I will consider it for a region $A$ in \emph{space} $\TTT^3$. Suppose the algebra $\alg(A)$ contains all spectral projections of all operators $Q(A')$ with $A'\subseteq A$. 
Define 
\be
\conf_A=\Gamma(A) \times \Gamma(A)
\ee
and note that
\be
\conf=\conf_A\times \conf_{A^c}
\ee
because, for $A\cap B=\emptyset$,
\be
\Gamma(A\cup B) = \Gamma(A) \times \Gamma(B)
\ee
in the sense that we identify $q\subset A\cup B$ with the pair $(q\cap A, q\cap B)$. Let $\mathcal{B}_A$ be the set of subsets of $\conf$ of the form $B\times \conf_{A^c}$ (i.e., the $\sigma$-algebra of events that depend only on goings-on in $A$, not in $A^c$). Then, and this is the relation between $P_\nat$ and $\alg(A)$, we should have that
\be\label{Pnatalg}
P_\nat(B) \in \alg(A)~~~\forall B\in\mathcal{B}_A\,.
\ee

\subsection{Spontaneous Pair Creation}

We now consider the free time evolution of the quantum Dirac field; let $H$ be the Hamiltonian, i.e., the positive second quantization of $H_1$. For definiteness, let 3-space be again $\TTT^3$.
There is a sense in which, according to $P_\nat$, electron-positron pairs can be created out of the vacuum; I refer to this fact as \emph{spontaneous pair creation}. 
In the literature, this expression usually means something else, viz., that in an external electromagnetic field $A_\mu(t,\vx)$ with $A_\mu(t=-\infty,\vx)=0=A_\mu(t=\infty,\vx)$, $\psi(t=-\infty)=\ket{\Omega}$ may evolve to $\psi(t=\infty)$ with $\psi^{11}\neq 0$ \cite[p.~362]{GR09}, \cite{PD08}.
Here, however, the expression means that
\be
[H, P_\nat(\Q^{n\nbar})]\neq 0.
\ee
That is, even if $A_\mu(t,\vx)=0$, for some states some probability gets transported to a different sector of $P_\nat$-particle number. 
Equivalently, the total electron and positron number operators are not conserved,
\be
[H,N_{\nat,\el}(\TTT^3)] \neq 0 \neq [H,N_{\nat,\pos}(\TTT^3)]\,,
\ee
in contrast to the relations
\be\label{Nobvconserved}
[H,N_{\obv,\el}(\TTT^3)]=0=[H,N_{\obv,\pos}(\TTT^3)]=[H,P_\obv(\Q^{n\nbar})]\,.
\ee

However, also according to $P_\nat$, the total charge operator is conserved,
\be
[H,Q(\TTT^3)]=0\,,
\ee
as follows, e.g., from \eqref{Nobvconserved} and \eqref{NobvQT3}. So the total charge sectors in configuration space,
\be
\mathcal{C}_z = \{q\in\Q:q_{\TTT^3}(q)=z\} = \bigcup_{\substack{n,\nbar=0\\\nbar-n=z}}^\infty \Q^{n\nbar}
\ee
for $z\in\ZZZ$, have zero exchange of probability between them; equivalently, the total charge sectors in Hilbert space,
\be
\text{range }P_\nat(\mathcal{C}_z)=
\bigoplus_{\substack{n,\nbar=0\\\nbar-n=z}}^\infty \Hilbert^{n\nbar}
\ee
for $z\in\ZZZ$, are invariant under $H$.

\subsection{Bohmian Trajectories}
\label{sec:Bohm}

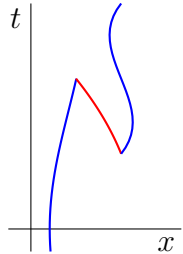
\begin{figure}[h]
\begin{center} 
  \begin{tikzpicture}
  \draw (-0.3,0) -- (2,0);
  \node at (1.8,-0.2) {$x$};
  \draw (0,-0.3) -- (0,3);
  \node at (-0.2,2.8) {$t$};
  \draw[thick,color=blue] plot[domain=-0.3:2] ({0.5-0.25*cos(\x r)},\x);
  \draw[thick,color=red] plot[domain=1:2] ({1.4-0.2*\x^2},\x);
  \draw[thick,color=blue] plot[domain=1:3] ({0.4*(\x-2)^3-0.4*\x+2},\x);
  \end{tikzpicture}
\end{center}
 \caption{Typical Bohmian trajectories: electron trajectories are shown in blue, positron in red. The particle number can change through pair creation or annihilation events. The trajectories are everywhere timelike or lightlike.}\label{fig:Bohm}
\end{figure}

The general principles for how to define the Bohmian motion in a configuration space $\Q$, given a Hilbert space $\Hilbert$, a Hamiltonian $H$, and a position POVM $P$ on $\conf$ acting on $\Hilbert$ \cite{crea2B} can be applied here with $P=P_\nat$. I briefly report that these principles lead to smooth, deterministic motion within one sector
\be
\Q^{n\nbar}=\bigl\{(q_\el,q_\pos)\in\Q:\#q_\el=n,\#q_\pos=\nbar\bigr\}
\ee
interrupted by stochastic jumps between the sectors with prescribed jump rates; an example of a realization is shown in Figure~\ref{fig:Bohm}.

In fact, these principles lead to the following definition of a stochastic process $Q_t$, the Bohm-Bell process, in $\Q$. $Q_0$ is distributed according to $\scp{\psi_0}{P_\nat(\cdot)|\psi_0}$. Between jumps of $Q_t=(Q_{t,\el},Q_{t,\pos})$, particles in $Q_{t,\el}=\{\vX_1(t),\ldots,\vX_n(t)\}$ move at velocity (no faster than light)
\be
\frac{d\vX_j^{\mu_j}}{dt} = \frac{J^{0...0\mu_j0...0}(t,Q_t)}{J^{0...0}(t,Q_t)}\ee
(and correspondingly for positrons) with
\begin{align}
&J^{\mu_1...\mu_n\mubar_1...\mubar_\nbar}(\vx_1...\vx_n,\vxbar_1...\vxbar_\nbar)=
\frac{1}{3^{n+\nbar}}\sum_{\substack{s_1...s'_1...\\ \sbar_1...\sbar'_1...}} (\gamma^0\gamma^{\mu_1})_{s_1s'_1}\cdots
(\gamma^0\gamma^{\mubar_1})_{\sbar'_1\sbar_1} \cdots\times\nonumber\\
&~~~~~~~~~~~~~~\times\langle \psi|\Psi^\dagger_{s_1}(\vx_1)\cdots\Psi^\dagger_{s_n}(\vx_n)\Psi_{\sbar_1}(\vxbar_1)\cdots \Psi_{\sbar_{\nbar}}(\vxbar_\nbar) \nonumber\\[3mm]
&~~~~~~\times P_\nat(\{\emptyset\})  \Psi^\dagger_{\sbar'_\nbar}(\vxbar_\nbar)\cdots \Psi^\dagger_{\sbar'_1}(\vxbar_1)\Psi_{s'_n}(\vx_n)\cdots\Psi_{s'_1}(\vx_1)| \psi\rangle.
\end{align}
This equation of motion is analogous to those based on Dirac wave functions in \cite{Bohm53} for a single particle, in \cite[Sec.s 10.5, 12.3]{BH} for many particles, or in \cite{HBD} for curved space-time or curved foliations. 

Electron-positron pairs spontaneously appear (emerging from the same point $\vx$), i.e., the configuration jumps $(q_\el,q_\pos) \to (q_\el\cup \vx,q_\pos\cup\vx)$, at rate (per $d^3\vx$) 
\be
\sigma=2\,\Im^+\frac{\scp{\psi}{P_\nat(q_\el\cup\vx, q_\pos\cup\vx) H P_\nat(q_\el,q_\pos)|\psi}}
{\scp{\psi}{P_\nat(q_\el,q_\pos)|\psi}}
\ee
with notation $s^+=\max\{x,0\}$ for $s\in\RRR$. (Here, I have slightly abused notation by denoting the density of $\scp{\psi}{P_\nat(dq)|\psi}$ relative to volume $dq$ by $\scp{\psi}{P_\nat(q)|\psi}$, etc.) Finally, when an electron and a positron meet, both disappear; this event often has positive probability although the configurations in which $q_\el\cap q_\pos \neq \emptyset$ form a set of measure zero in $\Q$. This completes the definition of the process.

A key property of the process is that at any time $t\in\RRR$, $Q_t$ has distribution $\scp{\psi_t}{P_\nat(\cdot)|\psi_t}$; this property is called the \emph{equivariance} of the distribution.

The perhaps easiest way to see that the Bohmian process $Q_t$ looks as described above is to consider the lattice approximation $\lattice$ as in Section~\ref{sec:discrete} with $P_{\nat,N}$, to define Bell's discrete jump process $\tilde Q_t$ with jump rates
\be
\tilde\sigma(q \to q') = 2\,\Im^+\frac{\scp{\psi}{P_{\nat,N}(q') H P_{\nat,N}(q)|\psi}}{\scp{\psi}{P_{\nat,N}(q)|\psi}}\,,
\ee
to consider the behavior of the pre-particles according to $\tilde Q_t$, and to use the known (though non-rigorous) fact \cite{Vink} that in the continuum limit jumps of particles to neighboring lattice sites associated with discretized differential operators become smooth motion according to Bohm's equation of motion. In the perspective of the pre-particles, $H$ conserves the number of particles and involves merely $H_1$ as a dicretized differential operator for each pre-particle, so each jump involves one pre-particle jumping to a neighboring lattice site, $\vx\to\vx'$. Denoting occupation numbers by $n_{\vx}$, the jump $2_{\vx} 1_{\vx'} \to 1_{\vx} 2_{\vx'}$ corresponds to, in the ontology associated with $P_\nat$, the jump of a positron from $\vx'$ to $\vx$ (which becomes smooth motion in the continuum limit). Likewise, the jump $3_{\vx} 2_{\vx'} \to 2_{\vx} 3_{\vx'}$ corresponds to the jump of an electron from $\vx$ to $\vx'$. The jump $2_{\vx} 2_{\vx'}\to 1_{\vx} 3_{\vx'}$ corresponds to the creation out of the vacuum of an electron at $\vx'$ and a positron at $\vx$, and the converse jump to pair annihilation, while occupation numbers 4 and 0 occur with probability 0 in the continuum limit.

The properties of the process $Q_t$ in $\Q$ include the local conservation of charge in every realization; the resulting global conservation of charge constitutes what was called a strong superselection rule in \cite{CDT05}. Moreover, the process is time reversal invariant in the sense that the distribution of the process $(Q_{-t}^\psi)_{t\in\RRR}$ in path space coincides with that of $(Q^{T\psi}_{t})_{t\in\RRR}$.

The Bohm-Bell process for $P_\obv$ was studied in \cite[Sec.~3.3]{crea2B}. It involves pair creation only in the presence of an external field $A_\mu$ and then has the striking property that when a pair is created, the electron and the positron do not pop up at the same location but at a distance comparable to the width of the projection of a 3d delta function to $\Hilbert_{1+}$ and thus to the Compton wave length of the electron. This property seems rather unphysical and can be regarded as another violation of interaction locality. 

Returning to $P_\nat$, the rate at which pairs get created depends on the quantum state. In the sea state $\ket{\Omega}$, this rate vanishes; in fact, eletrons and positrons have positive probability to be present (in equal number) but will not move; that is because $\ket{\Omega}$ is stationary and equals its own time reverse, so all currents vanish.

\subsection{Locality Properties}
\label{sec:loc}

There are several different locality properties. \emph{Locality} per se means that events at $x\in\RRR^4$ can't influence events at spacelike separated $y\in\RRR^4$. By Bell's theorem \cite{GNTZ}, this is violated in our universe.
\emph{Interaction locality (IL)} means that there is no interaction term in the Hamiltonian between spacelike separated regions. That seems to be the case in nature. IL appears equivalent to \emph{no superluminal signaling} (although the latter is perhaps less sharply defined because it refers to the possibilities of agents). \emph{Propagation locality (PL)} means that wave functions do not propagate faster than light. It seems to me that this is true in nature as well. In this section, I first give heuristic arguments to the effect that the free time evolution of the Dirac quantum field together with $P_\nat$ indeed satisfies both IL and PL; then I will comment on the possibility of precise criteria for IL and PL.

Here are three arguments for PL. First, the facts that the Bohmian particles move no faster than light and that the Bohmian configuration $Q_t$ is $\scp{\psi_t}{P_\nat(\cdot)|\psi_t}$ distributed at all times $t$ apparently entail PL. Second, in the lattice approximation, PL should hold in the ontology of Figure~\ref{fig:points3} if it holds in the ontology of Figure~\ref{fig:points2}; but in the latter situation, we are considering non-interacting particles governed by the free Dirac equation, so PL should hold; and the continuum limit should preserve PL. The third argument is a variant of the second based on a picture with an actual infinity of particles, using the Hilbert space $\Hilbert_\infty$ described in Appendix~\ref{app:Hilbertinfty} that allows for states with an infinite number of particles and contains $\Hilbert=\Fock(\Hilbert_{1+}\otimes \Fock(C\Hilbert_{1-})$ as a subspace. It seems that the action of $P_\nat$ can be extended to $\Hilbert_\infty$ if the configuration space is also extended so as to allow for configurations of infinitely many particles. Again, PL should hold in the ontology of Figure~\ref{fig:points3} if it holds in the ontology of Figure~\ref{fig:points2}; and the latter situation would apparently correspond in $\Hilbert_\infty$ to an infinite number of non-interacting particles governed by the free Dirac equation in continuous space, so PL should hold.

Here are three arguments for IL. First, the local algebras $\alg(\open_1)$ and $\alg(\open_2)$ should commute, assuming the space-time regions $\open_1,\open_2$ are spacelike separated. Since for every 3-set $A$ and time $t$ with $\{t\}\times A \subset \open_1$,
\be
e^{iHt}P_\nat(B)e^{-iHt}\in \alg(\open_1)
\ee
for all $B\in\mathcal{B}_A$ as in \eqref{Pnatalg}, all local operations in $\open_2$ should commute with $P_\nat(B)$ at time $t$, so there is no superluminal signaling between $\open_1$ and $\open_2$. Second, in the lattice approximation, IL should hold in the ontology of Figure~\ref{fig:points3} if it holds in the ontology of Figure~\ref{fig:points2}; but in the latter situation, we are considering non-interacting particles governed by the free Dirac equation, so IL should hold; and the continuum limit should preserve IL. Third, 
the same kind of argument can be done with $\Hilbert_\infty$ again.

Precise definitions of PL and IL were given in \cite{LT19}; however, they were intended for states near $|B\rangle$, not near $|\Omega\rangle$, and they are not appropriate in our setting. In detail, let the grown set $\mathrm{Gr}(A,t)$ and the shrunk set $\mathrm{Sr}(A,t)$ of $A\subseteq \RRR^3$ for $t>0$ be defined by 
\be
\mathrm{Gr}(A,t)= \bigcup_{\vx\in A} B_t(\vx) ~~\text{and}~~ \mathrm{Sr}(A,t) = \{\vx\in A: B_t(\vx)\subseteq A\}\,,
\ee
where $B_t(\vx)$ is the ball of radius $t$ around $\vx$. IL was defined in \cite{LT19} in terms of a factorization
\be\label{factorize}
\Hilbert=\Hilbert_A \otimes \Hilbert_{A^c}
\ee
for arbitrary 3-sets $A$ and their complements $A^c$; in our setting, $\Hilbert$ does not factorize in this way. PL was defined in \cite{LT19} in terms of a POVM $P$ on $\Gamma(\RRR^3)$ essentially by the condition
\be\label{PL19}
e^{-iHt}P(\emptyset(A))e^{iHt} \leq P(\emptyset(\mathrm{Sr}(A,|t|)))
\ee
with
\be\label{0Adef}
\emptyset(A)=\{q\in\Gamma(\RRR^3):q\cap A=\emptyset\}
\ee
the set of all configurations with no particles in $A$; in words, \eqref{PL19} expresses that if there were no particles in $A$ at time $0$, then there are no  particles in $\mathrm{Sr}(A,|t|)$ at time $t$. The problem is that this condition fails to separate PL from \emph{no particle creation from the vacuum (NCFV)}, and for $P_\nat$ we would not assume NCFV. As a replacement condition for PL near $|\Omega\rangle$, one might consider
\be
P(\emptyset(\mathrm{Gr}(A,|t|)\setminus \mathrm{Sr}(A,|t|)))
e^{-iHt}P(\emptyset(A))e^{iHt}P(\emptyset(\mathrm{Gr}(A,|t|)\setminus \mathrm{Sr}(A,|t|))) \leq P(q_A^{-1}(0))\,.
\ee
In words, this condition expresses that if there were no particles in $A$ at time $0$ and there are no particles in $\mathrm{Gr}(A,|t|)\setminus \mathrm{Sr}(A,|t|)$ at time $t$, then the charge in $A$ at time $t$ is 0 (i.e., there are as many electrons as positrons in $A$). This condition appears satisfied for $P_\nat$.

\subsection{Hegerfeldt-Malament Theorem}
\label{sec:Malament}

The Hegerfeldt-Malament theorem \cite{Heg74,HR80,Mal96,HC02} is often taken to exclude a particle ontology or position operators in relativistic quantum theory. So how does $P_\nat$ get around it? 
In a version relevant here, the theorem asserts:
{\it Suppose $\Hilbert$ is a Hilbert space (for 1 particle), the self-adjoint Hamiltonian $H$ is bounded from below, $P$ is a POVM on $\RRR^3$ acting on $\Hilbert$, and $U$ a unitary representation of the translation group of $\RRR^3$ on $\Hilbert$. Suppose further that (i) $P$ is translation covariant w.r.t.\ $U$,
\be
P(A+\va) = U_{\va} P(A) U_{\va}^{-1}~~~\forall \va\in\RRR^3\,,
\ee
and (ii) the following version of propagation locality holds:
\be\label{PL2}
e^{-iHt}P(A)e^{iHt}\leq P(\mathrm{Gr}(A,|t|))
\ee
with $\mathrm{Gr}(A,|t|)=\cup_{\vx\in A} B_{|t|}(\vx)$. Then a contradiction follows.}

Let us leave aside that $P_\nat$ on $\TTT^3$ may not fall under this theorem because the theorem assumes $\RRR^3$ as space. Then $P_\nat$ still violates two of the assumptions. First, on either $\TTT^3$ or $\RRR^3$, $P_\nat$ violates the assumption that the 1-particle sector, range $P(\conf^{10})$, is invariant under the time evolution. Second, on either $\TTT^3$ or $\RRR^3$, $P_\nat$ violates \eqref{PL2}; as discussed in Section~\ref{sec:loc}, in theories with particle creation, the condition \eqref{PL2}, which is equivalent to the 1-particle version of \eqref{PL19}, does not actually capture the idea that wave functions cannot propagate faster than light but includes as well that there is no creation of particles from the vacuum, which is violated by $P_\nat$. So, one cannot say that $P_\nat$ violates any requirements of relativity; rather, the assumptions of the Hegerfeldt-Malament theorem demand quite a bit more than what is required by relativity.

\bigskip

A similar theorem was proven already before Hegerfeldt and Malament by Borchers \cite[Thm.~III.1]{Bor67}: {\it Let $\varepsilon>0$. Suppose that on some Hilbert space $\Hilbert$, a self-adjoint Hamiltonian $H$ is bounded from below, and we have two projections $E,F$ such that (i)~$EF=0$ and (ii)~$F$ commutes with $e^{-iHt}Ee^{iHt}$ for all $|t|<\varepsilon$. Then $e^{-iHt}Ee^{iHt}F=0$ for all $t\in\RRR$.}

A difficulty for position operators seems to arise if $\Hilbert$ is a 1-particle Hilbert space and we have a position PVM $P$ on $\RRR^3$ or $\TTT^3$ acting on $\Hilbert$: let $E=P(A)$ for some bounded 3-region $A$ and $F=P(\mathrm{Gr}(A,\varepsilon)^c)$. If we assume propagation locality in the form \eqref{PL2}, then range $e^{-iHt}Ee^{iHt}$ is orthogonal to range $F$ for $|t|<\varepsilon$, and in particular the two operators commute. According to the theorem, range $e^{-iHt}Ee^{iHt}$ is then orthogonal to range $F$ for all $t$, which means that a state initially localized in $A$ will never overlap with $\mathrm{Gr}(A,\varepsilon)^c$, so, absurdly, states cannot propagate at all. Again, $P_\nat$ is not affected because the 1-particle sector is not invariant and \eqref{PL2} is not respected. 

Borchers' theorem can also be applied in another way, to position operators in the many-particle case: if $P$ is a PVM on $\Gamma(\RRR^3)$ or $\Gamma(\RRR^3)\times \Gamma(\RRR^3)$ (or $\Gamma(\TTT^3)$ or $\Gamma(\TTT^3)\times \Gamma(\TTT^3)$) acting on $\Hilbert$ (e.g., a Fock space or product thereof), let $E=P(\emptyset(A))$ (where $\emptyset(A)$ denotes, as in \eqref{0Adef}, the set of configurations with no particle in $A$) and $F=I-P(\emptyset(\mathrm{Sr}(A,\varepsilon)))$. Propagation locality in the form \eqref{PL19} would imply 
\be
e^{-iHt}Ee^{iHt} \leq P(\emptyset(\mathrm{Sr}(A,|t|))) \leq P(\emptyset(\mathrm{Sr}(A,\varepsilon))) =I-F
\ee
for $|t|<\varepsilon$. Thus, range $e^{-iHt}Ee^{iHt}$ is orthogonal to range $F$ for $|t|<\varepsilon$ and by the theorem for all $t$, which implies that if initially there are no particles in $A$ then, absurdly, no particle will ever reach $\mathrm{Sr}(A,\varepsilon)$. Again, $P_\nat$ is not affected because \eqref{PL19} is not respected.

\subsection{Reeh-Schlieder Theorem}

The Reeh-Schlieder theorem \cite{RS61} asserts:
{\it Let $\open\neq \emptyset$ be a bounded open region in Minkowski space-time. The set of all polynomials in $a^\dagger_t(P_{1+}x),a_t(P_{1+}x),b^\dagger_t(P_{1-}x),b_t(P_{1-}x)$ (Heisenberg-evolved) with $x=(\vx,s)\in\RRR^3\times\{1,2,3,4\}$ and $(t,\vx)\in \open$ applied to $\ket{\Omega}$ is dense in $\Hilbert$.}

That sounds paradoxical: Starting from the vacuum, we only act with operators localized in $\open$. We get any state, also with non-small particle probability at spacelike separation from $\open$.
But the sense of paradox evaporates when we accept that $\ket{\Omega}$ does not mean vacuum at all. Recall that
\be
P_\obv(\{\emptyset\})= \pr{\Omega} \neq P_\nat(\{\emptyset\})\,.
\ee
According to $P_\nat$, $\ket{\Omega}$ has a non-trivial distribution over configurations with non-zero particle number.

It may also be helpful to think of the discrete setting of the lattice $\lattice$ as in Section~\ref{sec:discrete}, where $\ket{\Omega}$ is an anti-symmetrized product of plane waves. Then $\ket\Omega$ looks like a highly entangled state, and it seems unsurprising that local operations on it can lead to all sorts of states.

It might also seem that the Reeh-Schlieder theorem contradicts the no-superluminal-signaling property, as it might seem to imply that operations in $\open$ can arrange, up to arbitrary precision, either state $\psi_1$ or state $\psi_2$ where $\psi_1$ and $\psi_2$ lead to very different probabilities for the outcomes of an experiment at spacelike separation from $\open$. However, not all polynomials in the $a^\dagger,a,b^\dagger,b$ at $(t,\vx)\in\open$ correspond to operations that can be carried out in $\open$. For comparison, from the Bell state $2^{-1/2}\bigl(\bigl|\uparrow\downarrow\bigr\rangle-\bigl|\downarrow\uparrow\bigr\rangle\bigr)$ it is possible to obtain all states in $\CCC^2 \otimes \CCC^2$ by applying operators of the form $S\otimes I$ with arbitrary operator $S:\CCC^2\to\CCC^2$ and $I$ the identity,\footnote{Proof: $\alpha \bigl|\uparrow\uparrow\bigr\rangle + \beta \bigl| \uparrow \downarrow \bigr\rangle + \gamma \bigl|\downarrow \uparrow\bigr\rangle + \delta \bigl|\downarrow \downarrow\bigr\rangle$ can be obtained with $S=2^{1/2}\bigl(-\alpha\bigl|\uparrow\bigr\rangle \bigl\langle\downarrow\bigr|+\beta \bigl|\uparrow\bigr\rangle \bigl\langle\uparrow\bigr| - \gamma \bigl|\downarrow\bigr\rangle \bigl\langle\downarrow\bigr| + \delta \bigl|\downarrow\bigr\rangle \bigl\langle\uparrow\bigr|\bigr)$.} despite the well-known no-signaling property for operations on just one side.

\subsection{What if Conjecture~\ref{conj} is False?} 

First, if space-time is discrete, then a version of $P_\nat$ resembling \eqref{PnatNdef} should still be well defined. Second, if space-time is continuous, then the $Q(A)$ operators would still be jointly diagonalizable while their joint eigenvalues $q(A)$ would no longer be $\sigma$-additive, only finitely additive. It may still be possible to take them seriously as an ontology. It would presumably be an ontology involving infinitely many positve and negative charges in every neighborhood of every $\vx\in\RRR^3$.

\section{Cauchy Surfaces and Curved Space-Time}
\label{sec:curved}

The construction of $P_\nat$ possesses a natural generalization to Cauchy surfaces (spacelike 3-surfaces, more or less) in space-time and to curved space-time, which I describe in this section. Since the Hamiltonian depends on the choice of space-time coordinates, it is not obvious which subspace $\Kilbert_1$ of $\Hilbert_1$ should play the role of $\Hilbert_{1-}$, the subspace ``to be filled by the Dirac sea.'' I will first assume that a choice of $\Kilbert_1$ is given for every Cauchy surface and later come back to question how $\Kilbert_1$ could be chosen, or whether we could avoid making such a choice.

\subsection{Construction of $\Hilbert_\Sigma$ and $P_{\nat\Sigma}$}

Let us consider a (time oriented, globally hyperbolic) Lorentzian 4-manifold $(\sM,g)$ as a given curved background space-time, and let us assume for convenience that Cauchy surfaces in $\sM$ are compact (have finite 3-volume like $\TTT^3$). We may also allow an external electromagnetic field $A_\mu$. It is well known how to define the 1-particle Dirac equation on $\sM$ (e.g., \cite{Dim82,DM14}): the 1-particle wave function $\varphi$ then is a cross-section of a rank-4 complex vector bundle $\sD$ over $\sM$ whose connection is fixed by the metric $g$ and $A_\mu$, and $\gamma$ is a cross-section of $\sD\otimes \sD' \otimes \CCC T\sM$ with $\sD'$ the dual bundle of $\sD$. For every Cauchy surface $\Sigma$, a Hilbert space $\Hilbert_{1\Sigma}$ is naturally defined as the set of cross-sections $\varphi$ of $\sD|_\Sigma$ such that $\scp{\varphi}{\varphi}_\Sigma < \infty$ for the inner product
\be\label{scp1Sigma}
\scp{\chi}{\varphi}_\Sigma := \int_\Sigma V(d^3x) \: \overline{\chi}(x) \, \gamma^\mu(x) \, n_\mu(x) \, \varphi(x)
\ee
with $V(d^3x) = \sqrt{|\det {}^3g|}\, d^3x$ the volume measure defined by the Riemann 3-metric ${}^3g$ on $\Sigma$ and $n_\mu(x)$ the future unit normal vector on $\Sigma$ at $x\in\Sigma$. The Dirac equation then defines a unitary 1-particle evolution $U_{1\Sigma}^{\Sigma'}:\Hilbert_{1\Sigma} \to \Hilbert_{1\Sigma'}$.

Now suppose we are given, for every Cauchy surface $\Sigma$, a ``sea space'' $\Kilbert_{1\Sigma} \subseteq \Hilbert_{1\Sigma}$ (also called a \emph{polarization} \cite{DDMS}). Then it is defined how to construct
\be\label{HilbertSigma}
\Hilbert_\Sigma=\Fock(\Kilbert_{1\Sigma}^\perp) \otimes \Fock(\overline{\Kilbert_{1\Sigma}})
\ee
along with $P_{\obv\Sigma}$ and $\rho_{\obv\Sigma}$ on $\conf_\Sigma=\Gamma(\Sigma)\times \Gamma(\Sigma)$. The definitions \eqref{abdef} of the annihilation and creation operators $a,a^\dagger,b,b^\dagger$ can be generalized in a straightforward way using the 1-particle inner product \eqref{scp1Sigma} (and elements of $\overline{\Kilbert_{1\Sigma}}$ rather than $C\Kilbert_{1\Sigma}$). We can thus also define field operators $\Psi(\vx):\Hilbert_\Sigma\to\Hilbert_\Sigma \otimes \sD_{\vx}$ for $\vx\in\Sigma$ as well as $\Psi^\dagger(\vx)$ and $\overline{\Psi}_s(\vx)= \sum_{s'} \Psi^\dagger_{s'}(\vx) (\gamma^\mu(\vx) n_\mu(\vx))_s^{s'}$ (in terms of a basis of $\sD_{\vx}$) and charge operators
\begin{subequations}\label{QAcurved}
\begin{align}
Q_\Sigma(A)&=-\int_A V(d^3x)\sum_s : \overline{\Psi}(x)\, \gamma^\mu(x) \, n_\mu (x)\,\Psi(x):\\
&=-\int_A V(d^3x)\sum_s  \Psi^\dagger(x)\, \Psi(x) +c
\end{align}
\end{subequations}
for $A\subseteq \Sigma$ with $c$ an infinite constant. It appears that the $Q_\Sigma(A)$ are self-adjoint and commute pairwise, so they can be simultaneously diagonalized. Whether the analog of Conjecture~\ref{conj} is true in this setting depends on the choice of $\Kilbert_{1\Sigma}$; if it is, $P_{\nat\Sigma}$ can be defined as the PVM on $\conf_\Sigma$ jointly diagonalizing all $Q_\Sigma(A)$.

It is known \cite{DDMS,DM16a,DM16b} that, as a generalization of the Shale-Stinespring \cite{SS} criterion, a unitary time evolution
\be\label{liftU}
U_{\Sigma}^{\Sigma'}: \Hilbert_\Sigma \to \Hilbert_{\Sigma'}
\ee
that lifts $U_{1\Sigma}^{\Sigma'}$ exists if and only if the sea spaces $\Kilbert_{1\Sigma}$ and $\Kilbert_{1\Sigma'}$ nearly coincide in terms of the 1-particle evolution, more precisely, if and only if 
\be\label{criterion}
P_{\Kilbert_{1\Sigma'}^\perp}U_{1\Sigma}^{\Sigma'}P_{\Kilbert_{1\Sigma}}
~\text{and}~
P_{\Kilbert_{1\Sigma'}}U_{1\Sigma}^{\Sigma'}P_{\Kilbert_{1\Sigma}^\perp}
\text{ are Hilbert-Schmidt operators.}
\ee
(This is the case, e.g., if $U_{1\Sigma}^{\Sigma'}\Kilbert_{1\Sigma}$ differs from $\Kilbert_{1\Sigma'}$ by only finitely many dimensions, i.e., if $U_{1\Sigma}^{\Sigma'}\Kilbert_{1\Sigma} \cap \Kilbert_{1\Sigma'}$ has finite codimension in both.) 

Anyway, if the analog of Conjecture~\ref{conj} holds for every $\Sigma$, and given a lift \eqref{liftU}, it appears that Bohmian trajectories can be defined for a given time foliation \cite{HBD} in the same way as in Section~\ref{sec:Bohm}, and IL and PL still hold in the appropriate sense. 

The probabilities with which detectors placed along $\Sigma$ find certain configurations $q$ may be expected to be given by $\scp{\psi_\Sigma}{P_{\nat\Sigma}(q)|\psi_\Sigma}$. However, these probabilities cannot be postulated independently for every $\Sigma$; they can be postulated (along with a collapse rule) at most for one foliation and then have to be proved for other Cauchy surfaces. Such proofs were provided in \cite{LT19,LT21} for evolution operators $U_\Sigma^{\Sigma'}$ and PVMs acting on Hilbert spaces $\Hilbert_\Sigma$ under the assumptions that there is no particle creation from the vacuum and $\Hilbert_\Sigma$ factorizes as in \eqref{factorize} (and, in \cite{LT19}, that the vacuum subspace is 1-dimensional), so these proofs do not apply to $P_\nat$. It would be of interest whether similar results can be proved for $P_\nat$.

\subsection{Choice of Sea Spaces}

Already in the 1-particle case and on a given Cauchy surface $\Sigma$, it is not defined what should be meant by the Hamiltonian on $\Sigma$; as becomes visible when we think of the Hamiltonian as representing more or less the unitary evolution to an infinitesimally neighboring Cauchy surface $\Sigma'$, the Hamiltonian depends not only on $\Sigma$ but also on $\Sigma'$ (and on a bijection between $\Sigma$ and $\Sigma'$) or, put differently, on the choice of lapse and shift functions. This situation keeps us from following the first instinct to take $\Kilbert_{1\Sigma}$ to be the negative spectral subspace of the Hamiltonian, and even makes such a choice implausible when a particularly obvious choice of $\Sigma$ and $\Sigma'$ (say, a particularly simple spacelike foliation) exists. This simple moral, that the negative spectral subspace of the Hamiltonian is not an option for the sea space, becomes clearly visible when considering a general curved space-time but is often not appreciated in the situation of an external electromagnetic field $A_\mu\neq 0$ in Minkowski space-time $\MMM^4$. 

In Minkowski space-time $\MMM^4$ with $A_\mu=0$, the situation is special as there exists an obvious choice of $\Kilbert_{1\Sigma}$ for every $\Sigma$ as follows. Any given $\varphi_\Sigma \in\Hilbert_{1\Sigma}$ uniquely defines a solution $\varphi:\MMM^4\to\CCC^4$ of the Dirac equation; the Fourier transform $\hat\varphi$ of $\varphi$ in all 4 variables is concentrated on the future mass shell and the past mass shell, so it can be decomposed (without requiring a choice of Lorentz frame) into a part $\hat\varphi_+$ on the future mass shell and a part $\hat\varphi_-$ on the past mass shell; Fourier transforming back, we obtain a decomposition $\varphi=\varphi_+ +\varphi_-$, and restricting to $\Sigma$ a decomposition $\varphi_\Sigma=\varphi_{\Sigma+}+\varphi_{\Sigma-}$. Since $\varphi_+$ and $\varphi_-$ are themselves solutions of the Dirac equation, we have that
\be
U_{1\Sigma}^{\Sigma'}\varphi_{\Sigma\pm}=\varphi_{\Sigma'\pm}\,. 
\ee
The fact that on $\Sigma_0=\{x^0=0\}$, in any Lorentz frame, the set of possible $\varphi_{\Sigma_0\pm}$ is just $\Hilbert_{1\pm}$ as in Section~\ref{sec:obv} shows that 
\be
\Hilbert_{1\Sigma\pm} := \{\varphi_{\Sigma\pm}:\varphi_{\Sigma}\in\Hilbert_{1\Sigma}\}
\ee
are two closed subspaces of $\Hilbert_{1\Sigma}$ that are mutually orthogonal. Set $\Kilbert_{1\Sigma} = \Hilbert_{1\Sigma-}$. Since $U_{1\Sigma}^{\Sigma'} \Kilbert_{1\Sigma} = \Kilbert_{1\Sigma'}$, this family of sea spaces $\Kilbert_{1\Sigma}$ satisfies the generalized Shale-Stinespring condition \eqref{criterion}. I conjecture that with these $\Kilbert_{1\Sigma}$, the analog of Conjecture~\ref{conj} holds on every Cauchy surface. We then obtain that for this natural choice of $\Kilbert_{1\Sigma}$ in Minkowski space-time, which is independent of any choice of Lorentz frame, $P_{\nat\Sigma}$ is defined on every $\Sigma$.

On a curved space-time $(\sM,g)$, Fourier transformation is not available, so the previous construction has no obvious analog. Here is an obvious choice of $\Kilbert_{1\Sigma}$ for general $\sM$: If on any $\Sigma_0$, $\Kilbert_{1\Sigma_0}$ were somehow given, we could transport it unitarily to every other $\Sigma$, i.e., we could set
\be\label{KilbertU}
\Kilbert_{1\Sigma} := U_{1\Sigma_0}^\Sigma \Kilbert_{1\Sigma_0}\,.
\ee
This choice would automatically satisfy the generalized Shale-Stinespring condition \eqref{criterion}. Actually, the fact that \eqref{criterion} demands that $\Kilbert_{1\Sigma}$ differs very little from $U_{1\Sigma_0}^\Sigma \Kilbert_{1\Sigma_0}$ may suggest this choice. However, this choice does not appear physically convincing. It implies that the evolution $U_\Sigma^{\Sigma'}$ on the full Hilbert space $\Hilbert_\Sigma$ always maps $|\Omega_\Sigma\rangle$ to $|\Omega_{\Sigma'}\rangle$, so that spontaneous pair creation in the traditional sense \cite{GR09,PD08} never occurs. In fact, suppose $\sM$ is asymptotically flat in the distant future and in the distant past and $A_\mu$ vanishes in the distant future and in the distant past, and on some flat surface $\Sigma_0$ in the past, $\Kilbert_{1\Sigma_0}$ is just the negative spectral subspace $\Hilbert_{1-}$ of the Hamiltonian of the local Lorentz frame associated with $\Sigma_0$. Then on a flat surface $\Sigma$ in the distant future, $\Kilbert_{1\Sigma}$ will in general not coincide with the negative spectral subspace of the Hamiltonian---the unitary 1-particle evolution does not bring back $\Kilbert_{1\Sigma}$ to its original place $\Kilbert_{1\Sigma_0}$.

If the laws of nature require the specification of $\Kilbert_{1\Sigma}$, then we need a law of nature specifying $\Kilbert_{1\Sigma}$. But no convincing candidate for such a law is in sight. So we may be inspired to try to avoid making any such choice.

\subsection{Without a Sea Space}

The definition of the standard Hilbert space \eqref{HilbertSigma} makes use of a sea space $\Kilbert_{1\Sigma}$, but the Hilbert space $\Hilbert_\infty$ of states involving infinitely many particles as defined in Appendix~\ref{app:Hilbertinfty} does not, and it contains a subspace $\Hilbert_\varphi$ that can be identified with the standard Hilbert space. This suggests that perhaps there is no fact in nature about which subspace $\Kilbert_{1\Sigma}$ of $\Hilbert_{1\Sigma}$ is the ``correct'' sea space, that it is a matter of initial conditions whether certain subspaces of $\Hilbert_{1\Sigma}$ happen to be ``filled,'' and that it is merely a matter of our terminology which 1-particle states we want to regard as electron states. Note, though, that in the ontology of $P_\nat$ there are facts in nature about which particles in a configuration are electrons and which are positrons.

To set up a theory of this form, we may apply the construction of $\Hilbert_\infty$ to each $\Hilbert_{1\Sigma}$ to obtain a Hilbert space $\Hilbert_{\infty\Sigma}$. Since field operators $\Psi_s(\vx)$ can be defined on $\Hilbert_\infty$, it appears that $P_\nat$ can perhaps also be defined on $\Hilbert_{\infty\Sigma}$ if we allow for a configuration space that includes configurations of infinitely many particles. The definition of the lifted non-interacting time evolution $U_\Sigma^{\Sigma'}:\Hilbert_{\infty\Sigma} \to \Hilbert_{\infty\Sigma'}$ is then straightforward, see Appendix~\ref{app:Hilbertinfty}, and should feature spontaneous pair creation in the traditional sense; this has to do with the fact that the standard Hilbert space $\Hilbert$ can be identified with different $\Hilbert_\varphi$'s. Another approach without a sea space is presented in \cite{positron2}.


\section{Conclusions}
\label{sec:last}

We have considered a set of position operators $P_\nat$ acting on the standard Hilbert space of the free Dirac quantum field; they form a PVM on the space of joint configurations of an arbitrary number of electrons and an arbitrary number of positrons. $P_\nat$ is different from the obvious position POVM $P_\obv$, and I have outlined reasons for thinking $P_\nat$ is more physically plausible than $P_\obv$. The most basic difference is perhaps that the sea state $|\Omega\rangle$, often called the vacuum state, does not correspond to particle vacuum according to $P_\nat$, although it does according to $P_\obv$. $P_\nat$ is, first, a candidate for representing detection probabilities; second, it can be used for defining a Bohm-type theory of electron and positron particles; and third, $P_\nat$ is a candidate for the missing link between quantum states of macroscopic objects and the probabilities of macroscopic states at times other than $-\infty$ and $+\infty$. I have also discussed the analogs of $P_\nat$ on a lattice and in curved space-time.

\appendix

\section{Conjugate Hilbert Space}
\label{app:conjugate}

The conjugate $\overline{\Kilbert}$ of a Hilbert space $\Kilbert$ is a Hilbert space that can be constructed as follows: $\overline\Kilbert$ has the same elements as $\Kilbert$ and the same addition, but a different rule of scalar multiplication: for any $\lambda\in\CCC$ and $v\in \overline\Kilbert=\Kilbert$,
\be
(\lambda,v) \mapsto \lambda^* v,
\ee
where the asterisk means complex conjugate. The inner product in $\overline\Kilbert$ is defined by
\be
(v,w) \mapsto \scp{v}{w}^*_\Kilbert
\ee
for any $v,w \in\overline{\Kilbert}=\Kilbert$; one easily verifies that it is conjugate-linear in $v$ and linear in $w$, and thus an inner product. The norm in $\overline\Kilbert$ agrees with that of $\Kilbert$, so $\overline\Kilbert$ inherits completeness from $\Kilbert$ and is a Hilbert space. Moreover, the identity mapping on $\Kilbert$ provides an anti-unitary anti-isomorphism $\Kilbert\to\overline\Kilbert$.

Alternatively and more abstractly, one can define the conjugate Hilbert space by a universal property as follows: $\overline\Kilbert$ is any Hilbert space equipped with an anti-unitary anti-isomorphism $J:\Kilbert\to\overline\Kilbert$. The fact that the pair $(\overline\Kilbert,J)$ is unique up to unitary isomorphism is exactly the ``universal property'': For any Hilbert space $\Hilbert$ and any anti-unitary anti-isomorphism $K:\Kilbert\to\Hilbert$, there exists a unitary isomorphism $U:\overline\Kilbert\to\Hilbert$ such that $K=UJ$. Indeed, if $J$ and $K$ are given, then $U=KJ^{-1}$ is a unitary isomorphism.

\section{POVMs}
\label{app:POVM}

In this appendix, we review basic facts about PVMs and POVMs and elucidate why, in the standard representation of vectors $\psi$ in $\Hilbert=\Fock(\Hilbert_{1+}) \otimes \Fock(C\Hilbert_{1-})$, the $|\psi|^2$ distribution corresponds to a POVM $P_\obv$ on $\Hilbert$.

\begin{defn}
A \emph{PVM (projection-valued measure)} on a measurable space $\conf$ acting on a Hilbert space $\Hilbert$ associates with every subset $B\subseteq \conf$ a projection $P(B)$ such that $P(\conf)=I$ and $P(B_1)+P(B_2)+\ldots= P(B_1 \cup B_2\cup \ldots)$ (``$\sigma$-additive'') whenever $B_i\cap B_j = \emptyset~\forall i\neq j$.
\end{defn}

For example, a PVM $P$ on $\conf$ acting on $\Hilbert=L^2(\conf,\sC)$ (with $\sC=\CCC$ or another Hilbert space) is defined by multiplication by the characteristic function of a set,
\be\label{obvPVM}
P(B)\psi(q) = 1_{B}(q)\, \psi(q) 
= \begin{cases} \psi(q) & \text{if } q\in B\\
 0 &\text{if } q\notin B.\end{cases}
\ee
It is the obvious PVM on any $L^2$ space. In non-relativistic quantum mechanics, this is the PVM on $\conf=\RRR^n$ acting on $\Hilbert=L^2(\RRR^n,\CCC^d)$ that corresponds to the position operators $X_i\psi(x_1\ldots x_n) = x_i\psi(x_1\ldots x_n)$.

\begin{defn}
A \emph{POVM (positive-operator-valued measure)} on a measurable space $\conf$ acting on a Hilbert space $\Hilbert$ associates with every subset $B\subseteq \conf$ a positive operator $P(B)$ such that $P(\conf)=I$ and $P(B_1)+P(B_2)+\ldots= P(B_1 \cup B_2\cup \ldots)$ whenever $B_i\cap B_j = \emptyset~\forall i\neq j$.
\end{defn}

So every PVM is a POVM but not vice versa. Note that a POVM $P$ and a unit vector $\psi\in\Hilbert$ always define a probability measure on $\conf$,
\be
\PPP(B) = \scp{\psi}{P(B)|\psi}.
\ee
In case $P$ is the obvious PVM on $\Hilbert=L^2(\conf,\sC)$ as in \eqref{obvPVM}, this probability distribution is the one with density $|\psi(q)|^2$.

As an example of a POVM that is not a PVM, if $\Hilbert$ is a closed subspace of a bigger Hilbert space $\Kilbert$, $P_\Hilbert$ the projection to $\Hilbert$, and $P$ a PVM on $\conf$ acting on $\Kilbert$, then
\be\label{projectedPOVM}
P'(B):=P_{\Hilbert} P(B) P_{\Hilbert}~~~\forall B \subseteq \conf
\ee
defines a POVM $P'$ on $\conf$ acting on $\Hilbert$. In particular, if $\Kilbert$ is an $L^2$ space and $P$ the obvious PVM as in \eqref{obvPVM}, then the POVM $P'$ corresponds to the $|\psi|^2$ distribution. This is the case with the standard representation of the quantized Dirac field (as described in Section~\ref{sec:intro}) and what I call the obvious POVM $P_\obv$: Since $\psi\in\Hilbert:=\Fock(\Hilbert_{1+})\otimes\Fock(C\Hilbert_{1-})$ can be regarded as a function on $\widehat\conf$ with values in (say) $\sC:=\bigoplus_{n,\nbar=0}^\infty (\CCC^4)^{\otimes (n+\nbar)}$, we can regard the Fock space $\Hilbert$ as a subspace of $\Kilbert:=L^2(\widehat\conf, \sC)$; thus, with $\pi$ the unordering map \eqref{unorderingdef},
\begin{align}
\PPP_\obv(B)
&=\int\limits_{\pi^{-1}(B)} \!\! dq \: \rho_\obv(q)\\
&=\int\limits_{\pi^{-1}(B)} \!\! dq \: |\psi(q)|^2\\
&=\scp{\psi}{P(\pi^{-1}(B))|\psi}_{\Kilbert}\\[2mm]
&=\scp{\psi}{P_\obv(B)|\psi}_{\Hilbert}
\end{align}
with 
\be
P_\obv(B) = P_\Hilbert 1_{\pi^{-1}(B)} P_\Hilbert 
\ee
in parallel to \eqref{projectedPOVM}.

\section{Hilbert Space of Infinitely Many Particles}
\label{app:Hilbertinfty}

A way of thinking in terms of infinitely many particles often suggests itself, with the sea state $|\Omega\rangle$ having infinitely more particles than the bottom state $|B\rangle$, and the ``top state'' $|T\rangle$ (with all energy levels of $H_1$ ``filled'') having infinitely more particles than $|\Omega\rangle$. This picture can be realized in an explicit construction of a Hilbert space $\Hilbert_\infty$ that contains $|B\rangle, |\Omega\rangle, |T\rangle$ and everything in between; both $\Fock(\Hilbert_1)$ and the standard Hilbert space $\Hilbert=\Fock(\Hilbert_{1+})\otimes \Fock(C\Hilbert_{1-})$ of the quantized Dirac field can be regarded as subspaces of $\Hilbert_\infty$. The construction is based on the concept of infinite wedge products $\varphi_1\wedge \varphi_2\wedge \ldots$ introduced by Deckert et al.~\cite{DDMS} (and independently and somewhat differently by Dimock \cite{Dim11}) and does not make use of the splitting $\Hilbert_1=\Hilbert_{1+} \oplus \Hilbert_{1-}$. Statements in this appendix are conjectures and have not been proved.

Let $S$ be the set of all finite or infinite sequences of elements of $\Hilbert_1$. Let $F=\CCC^S$ be the free complex vector space spanned by $S$. On $F$, define $B:F\times F\to \CCC$ as the unique sesquilinear form satisfying, for any $\chi=(\chi_1,\chi_2,\ldots)\in S$ and $\varphi=(\varphi_1,\varphi_2,\ldots)\in S$,
\be\label{Bdef}
B(\chi,\varphi) = \begin{cases} \det(\scp{\chi_i}{\varphi_j}_{\Hilbert_1})_{ij} & \text{if it exists}\\ 0&\text{otherwise.}\end{cases}
\ee
The determinant exists if and only if either both $\chi$ and $\varphi$ are finite sequences of the same length or both are infinite sequences and the infinite matrix $(\scp{\chi_i}{\varphi_j}_{\Hilbert_1})_{ij}$ possesses a Fredholm determinant. Let
\be
V=\{\varphi\in F: B(\varphi,\varphi)=0\}\,.
\ee
Then $V$ is a subspace of $F$, and $B$ defines an inner product on the quotient space $F/V$. Define $\Hilbert_\infty$ as the completion of $F/V$ with respect to this inner product.

For every $\varphi\in S$, we denote the equivalence class $\varphi+V$ by $\varphi_1\wedge \varphi_2\wedge \ldots$; the mapping $S\ni\varphi\mapsto \varphi+V\in\Hilbert_\infty$ turns out to be anti-symmetric and multi-linear, as expected of a wedge product.
The field operators $\Psi(f)$ and $\Psi^\dagger(f)$ for $f\in\Hilbert_1$ can be defined on $\Hilbert_\infty$ as follows. First, define $\Psi^\dagger(f)$ on $S$ as $(\varphi_1,\varphi_2,\ldots)\mapsto (f,\varphi_1,\varphi_2,\ldots)$, then on $F$ as the linear extension thereof; it turns out to be well defined on $F/V$ and on $\Hilbert_\infty$ by continuity. Define $\Psi(f)$ for $\varphi\in S$ with $\varphi_1=f$ by $\Psi(f)\varphi=\|f\|^2(\varphi_2,\varphi_3,\ldots)$, and for $\varphi\in S$ with $\scp{f}{\varphi_i}_{\Hilbert_1}=0$ for all $i$ by $\Psi(f)\varphi=0$; then $\Psi(f)$ possesses a unique linear extension to $F/V$ and by continuity to $\Hilbert_\infty$.

$\Hilbert_\infty$ is a non-separable Hilbert space, i.e., of uncountable dimension. $\Fock(\Hilbert_1)$ is naturally identified with the subspace ``finitely many particles away'' from $|B\rangle$, viz., the closed subspace of $\Hilbert_\infty$ spanned by the finite sequences. The standard Hilbert space $\Hilbert$ can be identified with a certain closed subspace of $\Hilbert_\infty$ as follows. Choose an ordered orthonormal basis $\varphi=(\varphi_1,\varphi_2,\ldots)$ of $\Hilbert_{1-}$, identify $|\Omega\rangle$ with $\varphi_1\wedge \varphi_2\wedge\ldots$, and form the closed subspace $\Hilbert_\varphi$ of $\Hilbert_\infty$ spanned by all $\chi_1\wedge \chi_2\wedge \ldots$ for sequences $(\chi_1,\chi_2,\ldots)$ obtained from $(\varphi_1,\varphi_2,\ldots)$ by removing finitely many vectors and including, anywhere in the sequence, finitely many new ones. Equivalently, $\Hilbert_\varphi$ is the closed subspace of $\Hilbert_\infty$ spanned by all vectors obtained through application of finitely many $\Psi(f)$ and $\Psi^\dagger(f)$ to $\varphi_1\wedge\varphi_2\wedge \ldots$. Note that different choices of the orthonormal basis $\varphi$ of the same subspace $\Hilbert_{1-}$ can lead to different, mutually orthogonal subspaces $\Hilbert_\varphi$ and thus to different identifications of $\Hilbert$ with a subspace of $\Hilbert_\infty$.

Every unitary $U_1:\Hilbert_1\to\Hilbert_1$ possesses an obvious lift $U_\infty:\Hilbert_\infty \to\Hilbert_\infty$ according to
\be\label{Uinfty}
U_\infty(\varphi_1\wedge \varphi_2 \wedge\ldots) =(U_1\varphi_1)\wedge (U_1\varphi_2)\wedge \ldots~.
\ee
On $\Hilbert_\varphi$, $U_\infty$ coincides up to a phase with the standard lift \cite[Sec.~10.3]{Tha} of $U_1$ to $\Hilbert$ whenever the latter exists, viz., whenever the Shale-Stinespring criterion \cite{SS} is satisfied, which demands that $P_{1+}U_1P_{1-}$ and $P_{1-}U_1P_{1+}$ are Hilbert-Schmidt operators.\footnote{Hilbert-Schmidt operators are those $S$ with $\tr S^2<\infty$.}

Likewise, for any $\Hilbert'_1 \neq \Hilbert_1$ and associated $\Hilbert'_\infty$ as above, every unitary isomorphism $U_1:\Hilbert_1\to\Hilbert'_1$ possesses an obvious lift $U_\infty:\Hilbert_\infty \to \Hilbert'_\infty$ given by the same formula \eqref{Uinfty}.

\bigskip

\noindent{\it Acknowledgments.} I wish to thank Christian Beck, Dirk Deckert, Shelly Goldstein, Sascha Lill, Markus N\"oth, Ward Struyve, Stefan Teufel, and Nino Zangh\`\i\ for helpful discussion.

\end{document}